\documentclass[twoside]{ima} 
\usepackage{amssymb,amsmath}
\begin{document}


\def\eqrefs#1#2{(\ref{#1}) and~(\ref{#2})}
\def\eqsref#1#2{(\ref{#1}) to~(\ref{#2})}
\def\sysref#1#2{(\ref{#1})--(\ref{#2})}

\def\Ref#1{Ref.~\cite{#1}}

\hyphenation{Eq Eqs Sec App Ref Fig}

\def\EQ{\begin{equation}}
\def\EQs{\begin{eqnarray}}
\def\endEQ{\end{equation}}
\def\endEQs{\end{eqnarray}}

\def\downupindices#1#2{{\mathstrut}^{}_{#1}{\mathstrut}_{}^{#2}}
\def\updownindices#1#2{{\mathstrut}_{}^{#1}{\mathstrut}^{}_{#2}}
\def\mixedindices#1#2{{\mathstrut}^{#1}_{#2}}
\def\downindex#1{{\mathstrut}_{#1}}
\def\upindex#1{{\mathstrut}^{#1}}

\def\eqtext#1{\hbox{\rm{#1}}}

\def\hp#1{\hphantom{#1}}

\def\frame#1#2{e\mixedindices{#2}{#1}}
\def\xframe#1{e_x\upindex{#1}}
\def\tframe#1{e_t\upindex{#1}}
\def\coframe#1{e\downindex{#1}}
\def\xcoframe#1{e_x\downindex{#1}}
\def\conx#1#2{\omega\downupindices{#1}{#2}}
\def\leftcoframe#1{\tilde e\downindex{#1}}
\def\leftconx#1#2{\tilde\omega\downupindices{#1}{#2}}
\def\xconx#1#2{\omega_x\downupindices{#1}{#2}}
\def\tconx#1#2{\omega_t\downupindices{#1}{#2}}
\def\gcovder#1{{}^g\nabla\downindex{#1}}
\def\covder#1{\nabla\downindex{#1}}
\def\v#1#2{v\mixedindices{#1}{#2}}
\def\bdsymb#1{{\boldsymbol{#1}}}
\def\curv#1#2{R\downupindices{#1}{#2}}
\def\w#1#2{\varpi\mixedindices{#1}{#2}}
\def\vs#1{{\mathfrak {#1}}}
\def\kconx{{}^{\vs{k}}\omega}
\def\D#1{D\downindex{#1}}
\def\Dinv#1{D\downindex{#1}^{-1}}
\def\trans#1{{#1}{}^T}
\def\htrans#1{{#1}{}^\dagger}
\def\vvec{\vec{v}}
\def\hvec{\vec{h}}
\def\wvec{\vec{\varpi}}
\def\Hop{{\mathcal H}}
\def\Jop{{\mathcal J}}
\def\Rop{{\mathcal R}}
\def\Iop{{\mathcal I}}
\def\Rnum#1{{\mathbb R}\upindex{#1}}
\def\Cnum#1{{\mathbb C}\upindex{#1}}
\def\idmatr{{\mathbb I}}
\def\h#1#2{h\mixedindices{#1}{#2}}
\def\hpar{h_\parallel}
\def\hperp{h_\perp}
\def\tr{{\rm tr}}
\def\ad{{\rm ad}}
\def\Ad{{\rm Ad}}
\def\covD#1{{\mathcal D}\downindex{#1}}
\def\hook{\lrcorner}
\def\i{{\rm i}}
\def\id#1#2{\delta\downupindices{#1}{#2}}
\def\c#1#2{c\downupindices{#1}{#2}}
\def\duc#1#2{c\updownindices{#1}{#2}}

\markboth{STEPHEN C. ANCO}{HAMILTONIAN CURVE FLOWS IN LIE GROUPS}

\title{HAMILTONIAN CURVE FLOWS IN LIE GROUPS\\ $G\subset U(N)$
AND VECTOR NLS, \lowercase{m}K\lowercase{d}V, SINE-GORDON SOLITON EQUATIONS}

\author{STEPHEN C. ANCO
\thanks{sanco@brocku.ca,
Department of Mathematics, Brock University, Canada}
}   

\maketitle   
             
\begin{abstract}                   
A bi-Hamiltonian hierarchy of complex vector soliton equations is derived
from geometric flows of non-stretching curves in the Lie groups
$G=SO(N+1)$, $SU(N)\subset U(N)$, generalizing previous work 
on integrable curve flows in Riemannian symmetric spaces $G/SO(N)$. 
The derivation uses a parallel frame and connection along the curves,
involving the Klein geometry of the group $G$. 
This is shown to yield 
the two known $U(N-1)$-invariant vector generalizations of
both the nonlinear Schr\"odinger (NLS) equation 
and the complex modified Korteweg-de Vries (mKdV) equation, 
as well as $U(N-1)$-invariant vector generalizations of 
the sine-Gordon (SG) equation found in recent symmetry-integrability
classifications of hyperbolic vector equations. 
The curve flows themselves are described in explicit form 
by chiral wave maps, chiral variants of Schr\"odinger maps,
and mKdV analogs. 
\end{abstract} 
 

{\AMSMOS 
37K05,37K10,37K25,35Q53,53C35
\endAMSMOS}

\section{Introduction}

The theory of integrable partial differential equations has many deep links
to the differential geometry of curves and surfaces. 
For instance 
the famous sine-Gordon (SG) and modified Korteveg-de Vries (mKdV)
soliton equations along with their common hierarchy of 
symmetries, conservation laws, and associated recursion operators
all can be encoded in geometric flows of non-stretching curves 
in Euclidean plane geometry \cite{GoldsteinPetrich,Nakayama}
by looking at the induced flow equation of the curvature invariant of 
such curves. 
A similar encoding is known to hold \cite{JPhysApaper}
in spherical geometry. 

Recent work \cite{kievpaper} has significantly generalized
this geometric origin of fundamental soliton equations to encompass
vector versions of mKdV and SG equations 
by considering non-stretching curve flows in Riemannian symmetric spaces
of the form $M=G/SO(N)$ for $N\ge 2$.
(Here $G$ represents the isometry group of $M$, 
and $SO(N)$ acts as a gauge group for the frame bundle of $M$,
such that $SO(N) \subset G$ is an invariant subgroup under
an involutive automorphism of $G$.)
Such spaces \cite{Helgason} are exhausted by the groups
$G=SO(N+1),SU(N)$
and describe curved $G$-invariant geometries
that are a natural generalization of Euclidean spaces.
In particular, for $N=2$, the local isomorphism $SO(3)\simeq SU(2)$ implies
both of these spaces are isometric to the standard 2-sphere geometry, 
$S^2 \simeq G/SO(2)$. 

As main results in \cite{kievpaper},
it was shown firstly that there is a geometric encoding of 
$O(N-1)$-invariant bi-Hamiltonian operators in 
the Cartan structure equations for torsion and curvature of 
a moving parallel frame and its associated frame connection $1$-form
for non-stretching curves in the spaces $G/SO(N)$ 
viewed as Klein geometries \cite{Sharpe}.
The group $O(N-1)$ here arises as the isotropy subgroup 
in the gauge group $SO(N)$ preserving the parallel property of 
the moving frame. 
Secondly, this bi-Hamiltonian structure generates 
a hierarchy of integrable flows of curves in which the frame components of the
principal normal along the curve satisfy 
$O(N-1)$-invariant vector soliton equations 
related by a hereditary recursion operator. 
These normal components in a parallel moving frame have
the geometrical meaning of curvature covariants of curves
relative to the isotropy group $O(N-1)$. 
Thirdly, the two isometry groups $G=SO(N+1),SU(N)$ were shown to 
give different hierarchies whose 
$O(N-1)$-invariant vector evolution equations of lowest-order 
are precisely the two known vector versions of integrable mKdV equations 
found in the symmetry-integrability classifications presented 
in \cite{SokolovWolf}.
In addition these hierarchies were shown to also contain
$O(N-1)$-invariant vector hyperbolic equations given by 
two different vector versions of integrable SG equations that are known 
from a recent generalization of symmetry-integrability classifications
to the hyperbolic case \cite{AncoWolf}.
Finally, 
the geometric curve flows corresponding to these vector SG and mKdV equations 
in both hierarchies were found to be described by 
wave maps and mKdV analogs of Schrodinger maps
into the curved spaces $SO(N+1)/SO(N)$, $SU(N)/SO(N)$.

The present paper extends the same analysis to give 
a geometric origin for $U(N-1)$-invariant vector soliton equations
and their bi-Hamiltonian integrability structure
from considering flows of non-stretching curves in the Lie groups
$G=SO(N+1),SU(N) \subset U(N)$. 
A main idea will be to view these Lie groups as Klein geometries
carrying the structure of a Riemannian symmetric space \cite{Helgason}
given by $G\simeq G\times G/\diag(G\times G) =M$ for $N\ge 2$
(with $G$ thus representing both the isometry group of $M$
as well as the gauge group for the frame bundle of $M$). 
Note for $N=2$ both these spaces locally describe a 3-sphere geometry,
$S^3 \simeq SO(3)\simeq SU(2)$.

In this setting the parallel moving frame formulation of
non-stretching curves developed in \cite{kievpaper} for
the Riemannian symmetric spaces 
\break $SO(N+1)/SO(N)$, $SU(N)/SO(N)$
can be applied directly to the Lie groups $SO(N+1)$, $SU(N)$ themselves,
where the isotropy subgroup 
\break $O(N-1)\subset SO(N)$ of such frames 
is replaced by $U(N-1)\subset SU(N)$ and $U(1)\times O(N-1)\subset SO(N+1)$ 
in the two respective cases. 
As a result,
it will be shown that the frame structure equations 
geometrically encode $U(N-1)$-invariant bi-Hamiltonian operators
that generate a hierarchy of integrable flows of curves
in both spaces $G=SO(N+1),SU(N)$. 
Moreover, the frame components of the 
principal normal along the curves in the two hierarchies
will be seen to satisfy 
$U(N-1)$-invariant vector soliton equations that exhaust
the two known vector versions of integrable NLS equations
and corresponding complex vector versions of integrable mKdV equations, 
as well as the two known complex vector versions of integrable SG equations,
found respectively in the symmetry-integrability classifications 
stated in \cite{SokolovWolf} and \cite{AncoWolf}. 
Lastly, the geometric curve flows arising from these 
vector SG, NLS, and mKdV equations in both hierarchies
will be shown to consist of chiral wave maps,
chiral variants of Schrodinger maps and their mKdV analogs, 
into the curved spaces $SO(N+1)$, $SU(N)$.

Taken together, the results here and in \cite{kievpaper}
geometrically account for the existence and the bi-Hamiltonian 
integrability structure of all known 
vector generalizations of NLS, mKdV, SG soliton equations. 

Related work \cite{ChouQu1,ChouQu2,ChouQu3,ChouQu4} 
has obtained geometric derivations of the KdV equation and other 
scalar soliton equations along with their hereditary integrability structure
from non-stretching curve flows in planar Klein geometries
(which are group-theoretic generalizations of the Euclidean plane
such that the Euclidean group is replaced by a different isometry group
acting locally and effectively on $\Rnum{2}$).

Previous results on integrable vector NLS and mKdV equations geometrically
associated to Lie groups in the Riemannian case appeared in 
\cite{FordyKulish,Fordy,AthorneFordy,Athorne,JPhysApaper}.
Earlier work on deriving vector SG equations 
from Riemannian symmetric spaces and Lie groups 
can be found in 
\cite{Bakas,Pohlmeyer,JPhysApaper}. 
The basic idea of studying curve flows via parallel moving frames 
appears in \cite{LangerPerline,SandersWang1,SandersWang2}.

\section{Curve flows and parallel frames}

Compact semisimple Lie groups $G$ are well-known to have 
the natural structure of a Riemannian manifold
whose metric tensor $g$ arises in a left-invariant fashion 
\cite{KobayashiNomizu}
from the Cartan-Killing inner product $<\cdot,\cdot>_\vs{g}$
on the Lie algebra $\vs{g}$ of $G$.
This structure can be formulated in an explicit way by the introduction of
a left-invariant orthonormal frame $\frame{a}{}$ on $G$,
satisfying the commutator property 
$[\frame{a}{},\frame{b}{}]=\c{ab}{c}\frame{c}{}$
where $\c{ab}{c}$ denotes the structure constants of $\vs{g}$,
and frame indices $a,b$, etc. run $1,\ldots,n$
where $n=\dim G$. 
The Riemannian metric tensor $g$ on $G$ is then given by 
\EQ
g(\frame{a}{},\frame{b}{})=-\frac{1}{2}\c{ac}{d}\c{bd}{c} = \id{ab}{}
\endEQ
while
\EQ
\curv{c}{d}(\frame{a}{},\frame{b}{})=\c{ab}{f}\c{cf}{d}
\endEQ
yields the Riemannian curvature tensor of $G$ expressed as a linear map
$[\gcovder{},\gcovder{}] =\curv{}{}(\cdot,\cdot)$.
The frame vectors $\frame{a}{}$ also determine 
connection 1-forms 
\EQ\label{invframe}
\conx{}{ab} = \duc{ab}{c} \frame{}{c}
\endEQ
in terms of coframe 1-forms $\frame{}{a}$ dual to $\frame{a}{}$
obeying the standard frame structure equations \cite{KobayashiNomizu}
\EQs
&& \gcovder{}\frame{}{a} = \conx{b}{a}\otimes\frame{}{b} ,
\\
&& [\gcovder{},\gcovder{}] \frame{}{a} = \curv{b}{a}(\cdot,\cdot) \frame{}{b} ,
\endEQs
with
\EQ
\curv{b}{a}(\cdot,\cdot) = d\conx{b}{a}+\conx{b}{c}\wedge\conx{c}{a}
\endEQ
where $d$ denotes the exterior total derivative on $G$
and $\gcovder{}$ denotes the Riemannian covariant derivative on $G$.
Note that frame indices are raised and lowered using 
$\id{ab}{}= \diag(+1,\ldots,+1)$,
and the summation convention is used for repeated indices. 

Now let $\gamma(t,x)$ be a flow of a non-stretching curve in $G$. 
Write $Y=\gamma_{t}$ for the evolution vector of the curve and write
$X=\gamma_{x}$ for the tangent vector along the curve normalized by
$g(X,X)=1$, which is the condition for $\gamma$ to be non-stretching,
so thus $x$ represents arclength.
Suppose $\frame{a}{}$ is a moving parallel frame \cite{Bishop}
along the curve $\gamma$.
Specifically, in the two-dimensional tangent space $T_\gamma M$ of the flow,
$\frame{a}{}$ is assumed to be adapted to $\gamma$ via
\EQ
\frame{}{a} := X\ (a=1),\quad (\frame{}{a})_{\perp}\ (a=2,\ldots,n)
\endEQ
where $g(X,(\frame{}{a})_{\perp})=0$, such that the covariant
derivative of each of the $n-1$ normal vectors $(\frame{}{a})_\perp$
in the frame is tangent to $\gamma$,
\EQ\label{parallelconxtang}
\gcovder{x} (\frame{}{a})_{\perp} =-\v{a}{} X
\endEQ
holding for some functions $\v{a}{}$,
while the covariant derivative of the tangent vector
$X$ in the frame is normal to $\gamma$,
\EQ\label{parallelconxperp}
\gcovder{x} X =
\v{a}{}(\frame{a}{})_{\perp} .
\endEQ
Equivalently, in the notation of \cite{kievpaper}, 
the components of the connection $1$-forms of the parallel frame
along $\gamma$ 
are given by the skew matrix
\EQ\label{riemannconx}
\xconx{a}{b}:=X \hook \conx{a}{b} = \xcoframe{a} \v{b}{} - \xframe{b} \v{}{a}
= 
\begin{pmatrix}
0 & \v{b}{}\\ -\v{}{a}&\bdsymb{0}
\end{pmatrix} 
\endEQ
where 
\EQ\label{riemannframe}
\xframe{a}:=g(X,\frame{}{a})
= (1, \vec{0}) 
\endEQ
is the row matrix of the frame
in the tangent direction.
(Throughout, upper/lower frame indices will represent row/column matrices.)
This description gives a purely algebraic characterization \cite{kievpaper}
of a parallel frame:
$\xframe{a}$ is a fixed unit vector in $\Rnum{n}\simeq T_x G$
preserved by a $SO(n-1)$ rotation subgroup of
the local frame structure group $SO(n)$
with $G$ viewed as being a $n$-dimensional Riemannian manifold,
while $\xconx{a}{b}$ belongs to the orthogonal complement of
the corresponding rotation subalgebra $\vs{so}(n-1)$
in the Lie algebra $\vs{so}(n)$ of $SO(n)$.

However, taking into account the left-invariance property \eqref{invframe},
note $\xconx{}{ab} = \duc{ab}{c}\xframe{c}$
and consequently  $\duc{ab}{c} = 2\id{c}{[a} \v{b]}{}$
which implies degeneracy of the structure constants,
namely $\duc{ab}{c} \v{c}{}=0$ and $\c{[abc]}{}=0$.
But such conditions are impossible 
in a non-abelian semisimple Lie algebra \cite{liealgebra},
and hence there do not exist parallel frames that are left-invariant. 
This difficulty can be by-passed if 
moving parallel frames are introduced in a setting that uses
the structure of $G$ as a Klein geometry
rather than a left-invariant Riemannian manifold,
which will relax the precondition for parallel frames to be left-invariant. 

The Klein geometry of a compact semisimple Lie group $G$ is given by 
\cite{Sharpe,KobayashiNomizu}
the Riemannian symmetric space $M=K/H \simeq G$
for $K=G\times G \supset H=\diag K \simeq G$
in which $K$ is regarded as a principle $G$ bundle over $M$.
Note $H$ is a Lie subgroup of $K$ invariant under an involutive automorphism
$\sigma$ given by a permutation of the factors $G$ in $K$.
The Riemannian structure of $M$ is isomorphic with that of $G$ itself.
In particular, 
under the canonical mapping of $G$ into $K/H \simeq G$, 
the Riemannian curvature tensor and metric tensor on $M$
pull back to 
the standard ones $\curv{}{}(\cdot,\cdot)$ and $g$ on $G$,
both of which are covariantly constant and $G$-invariant. 
The primary difference in regarding $G$ as a Klein geometry is that
its frame bundle \cite{Sharpe} 
will naturally have $G$ for the gauge group,
which is a reduction of the $SO(n)$ frame bundle of $G$
as a purely Riemannian manifold.
\footnote{More details will be given elsewhere \cite{forthcoming}.}

In the same manner as for the Klein geometries considered in \cite{kievpaper},
the frame structure equations for non-stretching curve flows in 
the space $M= K/H \simeq G$ can be shown to 
directly encode a bi-Hamiltonian structure based on geometrical variables,
utilizing a moving parallel frame combined with the property of 
the geometry of $M$ that its frame curvature matrix is constant on $M$.
In addition the resulting bi-Hamiltonian structure 
will be invariant under the isotropy subgroup of $H$ that 
preserves the parallel property of the frame thereby leaving $X$ invariant. 
Since in the present work we are seeking 
$U(N-1)$-invariant bi-Hamiltonian operators,
the simplest situation will be to have 
$U(N-1) \subset SU(N) =H$. 
Hence we first will consider the Klein geometry 
$M=K/H \simeq SU(N)$ given by the Lie group $G=SU(N)$. 

The frame bundle structure of this space $M=K/H \simeq SU(N)$ is tied to 
a zero-curvature connection $1$-form $\omega_K$ 
called the Cartan connection \cite{Sharpe}
which is identified with 
the left-invariant $\vs{k}$-valued Maurer-Cartan form on the
Lie group $K=SU(N)\times SU(N)$.
Here $\vs{k} =\vs{su}(N)\oplus\vs{su}(N)$ is the Lie algebra of $K$
and $\vs{h}=\diag \vs{k} \simeq \vs{su}(N)$ 
is the Lie subalgebra in $\vs{k}$
corresponding to the gauge group 
$H=\diag(SU(N)\times SU(N)) \simeq SU(N)$ in $K$.
The involutive automorphism $\sigma$ of $K$ induces the decomposition
$\vs{k}=\vs{p}\oplus\vs{h}$ 
where $\vs{h}$ and $\vs{p}$ are respective eigenspaces 
$\sigma =\pm 1$ in $\vs{k}$,
where $\sigma$ permutes the $\vs{su}(N)$ factors of $\vs{k}$.
The subspace $\vs{p}\simeq \vs{su}(N)$ has the Lie bracket relations
\EQ
[\vs{p},\vs{p}] \subset \vs{h} \simeq \vs{su}(N) ,\quad
[\vs{h},\vs{p}] \simeq [\vs{su}(N),\vs{p}] \subset \vs{p} .
\endEQ
In particular there is a natural action of 
$\vs{h}\simeq\vs{su}(N)$ on $\vs{p}$. 
To proceed, 
in the group $K$ 
regarded as a principal $SU(N)$ bundle over $M$,
fix any local section 
and pull-back $\omega_K$ to give a $\vs{k}$-valued $1$-form $\kconx{}$
at $x$ in $M$.
The effect of changing the local section is to induce
a $SU(N)$ gauge transformation on $\kconx{}$.
We now decompose $\kconx{}$ with respect to $\sigma$. 
It can be shown that \cite{Sharpe} the symmetric part
\EQ\label{kleinconx}
\conx{}{}:= \frac{1}{2}\kconx{} + \frac{1}{2}\sigma(\kconx{})
\endEQ
defines a $\vs{su}(N)$-valued connection $1$-form for the group
action of $SU(N)$ on the tangent space $T_x M \simeq \vs{p}$,
while the antisymmetric part
\EQ\label{kleinframe}
\coframe{}:=\frac{1}{2}\kconx{} - \frac{1}{2}\sigma(\kconx{})
\endEQ
defines a $\vs{p}$-valued coframe for the Cartan-Killing
inner product $<\cdot,\cdot>_{\vs{p}}$ on $T_x K \simeq \vs{k}$
restricted to $T_x M \simeq \vs{p}$.
This inner product 
\footnote{The sign convention that $<\cdot,\cdot>_\vs{p}$ is positive-definite
will be used for convenience.}
provides a Riemannian metric
\EQ
g=<\coframe{} \otimes \coframe{}>_\vs{p}
\endEQ
on $M\simeq SU(N)$,
such that the squared norm of any vector $X\in T_x M$ is given by
$|X|_g^2 = g(X,X)=<X\hook \coframe{},X\hook \coframe{}>_\vs{p}$.
Note $\coframe{}$ and $\conx{}{}$ will be left-invariant
with respect to the group action of $H\simeq SU(N)$ if and only if 
the local section of the $SU(N)$ bundle $K$ used to define
the 1-form $\kconx{}$ is a left-invariant function. 
In particular, 
if $h:M\rightarrow H$ is an $SU(N)$ gauge transformation
relating a left-invariant section to an arbitrary local section of $K$
then 
$\leftcoframe{} =\Ad(h^{-1})\coframe{}$
and 
$\leftconx{}{} =\Ad(h^{-1})\conx{}{} +h^{-1}dh$
will be a left-invariant coframe and connection on $M\simeq SU(N)$. 

Moreover, associated to this structure provided by the Maurer-Cartan form, 
there is a $SU(N)$-invariant covariant derivative $\covder{}$
whose restriction to the tangent space $T_\gamma M$
for any curve flow $\gamma(t,x)$ in $M\simeq SU(N)$ is defined
via
\EQ\label{ewrelation}
\covder{x} \coframe{} =
[\coframe{},\gamma_{x}\hook \conx{}{}]
\qquad\eqtext{ and }\qquad
\covder{t} \coframe{} =
[\coframe{},\gamma_{t} \hook\conx{}{}] .
\endEQ
These covariant derivatives obey the Cartan structure equations 
obtained from a decomposition of
the zero-curvature equation of the Maurer-Cartan form 
\EQ
0 = d\omega_K + \frac{1}{2} [\omega_K,\omega_K] .
\endEQ
Namely $\covder{x}, \covder{t}$ have zero torsion
\EQ\label{cartantors}
0 = (\covder{x} \gamma_{t} - \covder{t} \gamma_{x})\hook \coframe{} =
\D{x}\tframe{} - \D{t}\xframe{}
+ [\xconx{}{},\tframe{}] -[\tconx{}{},\xframe{}]
\endEQ
and carry $SU(N)$-invariant curvature
\EQs
\label{cartancurv}
\curv{}{}(\gamma_{x},\gamma_{t}) \coframe{}
&=&
[\covder{x},\covder{t}] \coframe{}
= \D{x} \tconx{}{} - \D{t} \xconx{}{} + [\xconx{}{},\tconx{}{}]
\\ &=&
-[\xframe{},\tframe{}]
\nonumber
\endEQs
where
\EQ
\xframe{}:= \gamma_{x} \hook \coframe{} ,\qquad
\tframe{}:= \gamma_{t} \hook \coframe{} ,\qquad
\xconx{}{}:= \gamma_{x} \hook \conx{}{} ,\qquad
\tconx{}{}:= \gamma_{t} \hook \conx{}{} .
\endEQ

\begin{remark}
The soldering relations \eqrefs{kleinconx}{kleinframe} 
together with the canonical identifications 
$\vs{p}\simeq\vs{su}(N)$ and $\vs{h}\simeq\vs{su}(N)$
provide an isomorphism between 
the Klein geometry of $M\simeq SU(N)$ 
and the Riemannian geometry of the Lie group $G=SU(N)$. 
This isomorphism allows $\frame{}{}$ and $\conx{}{}$
to be regarded hereafter as an $\vs{su}(N)$-valued coframe 
and its associated $\vs{su}(N)$-valued connection 1-form 
introduced on $G=SU(N)$ itself,
without the property of left-invariance. 
\end{remark}

Geometrically, it thus follows that 
$\xframe{}$ and $\xconx{}{}$ represent the tangential part of
the coframe and the connection $1$-form along $\gamma$.
For a non-stretching curve $\gamma$,
where $x$ is the arclength,
note $\xframe{}$ has unit norm in the inner product,
$<\xframe{},\xframe{}>_{\vs{p}}=1$.
This implies $\vs{p}\simeq\vs{su}(N)$ has a decomposition into
tangential and normal subspaces
$\vs{p}_{\parallel}$ and $\vs{p}_{\perp}$ with respect
to $\xframe{}$ such that $<\xframe{},\vs{p}_{\perp}>_{\vs{p}}=0$,
with $\vs{p}=\vs{p}_{\perp} \oplus \vs{p}_{\parallel} \simeq\vs{su}(N)$ 
and $\vs{p}_{\parallel} \simeq \Rnum{}$.

The isotropy subgroup in $H\simeq SU(N)$ preserving $\xframe{}$
is clearly the unitary group $U(N-1) \subset SU(N)$ acting on 
$\vs{p}\simeq \vs{su}(N)$.
This motivates an algebraic characterization of 
a parallel frame \cite{kievpaper}
defined by the property  that 
$\xconx{}{}$ belongs to the orthogonal complement of
the $U(N-1)$ unitary rotation Lie subalgebra $\vs{u}(N-1)$
contained in the Lie algebra $\vs{su}(N)$ of $SU(N)$,
in analogy to the Riemannian case. 
Its geometrical meaning, however,
generalizes the Riemannian properties
\eqrefs{parallelconxtang}{parallelconxperp},
as follows.
Let $\frame{a}{}$ be a frame whose dual coframe
is identified
with the $\vs{su}(N)$-valued coframe $\coframe{}$
in a fixed orthonormal basis for $\vs{p}\simeq\vs{su}(N)$.
Decompose the coframe into parallel/perpendicular parts
with respect to $\xframe{}$ in an algebraic sense
as defined by
the kernel/cokernel of Lie algebra multiplication
$[\xframe{},\cdot\ ]_\vs{p}=\ad(\xframe{})$.
Thus we have
$\coframe{}=(\coframe{C},\coframe{C^\perp})$
where the $\vs{su}(N)$-valued covectors $\coframe{C},\coframe{C^\perp}$ 
satisfy 
$[\xframe{},\coframe{C}]_\vs{p}=0$,
and
$\coframe{C^\perp}$ is orthogonal to $\coframe{C}$,
so
$[\xframe{},\coframe{C^\perp}]_\vs{p} \neq0$.
Note there is a corresponding algebraic decomposition of
the tangent space $T_x G \simeq \vs{su}(N)\simeq \vs{p}$ given by
$\vs{p}=\vs{p}_C \oplus \vs{p}_{C^\perp}$, with
$\vs{p}_\parallel \subseteq \vs{p}_C$
and
$\vs{p}_{C^\perp} \subseteq \vs{p}_\perp$,
where $[\vs{p}_\parallel,\vs{p}_C]=0$
and $<\vs{p}_{C^\perp},\vs{p}_C>_\vs{p}=0$,
so $[\vs{p}_\parallel,\vs{p}_{C^\perp}]\neq 0$
(namely, $\vs{p}_C$ is the centralizer of $\xframe{}$
in $\vs{p} \simeq \vs{su}(N)$).
This decomposition is preserved by $\ad(\xconx{}{})$
which acts as an infinitesimal unitary rotation,
$\ad(\xconx{}{})\vs{p}_C \subseteq \vs{p}_{C^\perp}$,
$\ad(\xconx{}{})\vs{p}_{C^\perp} \subseteq \vs{p}_C$.
Hence, from equation \eqref{ewrelation},
the derivative $\covder{x}$ of a covector perpendicular
(respectively parallel) to $\xframe{}$
lies parallel (respectively perpendicular) to $\xframe{}$,
namely
$\covder{x}\coframe{C}$ belongs to $\vs{p}_{C^\perp}$,
$\covder{x}\coframe{C^\perp}$ belongs to $\vs{p}_C$.
In the Riemannian setting,
these properties correspond to
$\gcovder{x}(\frame{}{a})_C = \v{a}{\ b}(\frame{}{b})_{C^\perp}$,
$\gcovder{x}(\frame{}{a})_{C^\perp} = -\v{\ a}{b}(\frame{}{b})_C$
for some functions $\v{ab}{}=-\v{ba}{}$,
without the left-invariance property \eqref{invframe}.
Such a frame will be called {\it $SU(N)$-parallel}
and defines a strict generalization of a Riemannian parallel frame
whenever $\vs{p}_C$ is larger than $\vs{p}_\parallel$.

It should be noted that the existence of a $SU(N)$-parallel frame 
for curve flows in the Klein geometry $M=K/H \simeq SU(N)$ 
is guaranteed by the $SU(N)$ gauge freedom
on $\frame{}{}$ and $\conx{}{}$ inherited from
the local section of $K=SU(N)\times SU(N)$ 
used to pull back the Maurer-Cartan form
to $M$.

All these developments carry over to the Lie group $G=SO(N+1)$
viewed as a Klein geometry $M=K/H \simeq SO(N+1)$
for $K=$ 
\break $SO(N+1)\times SO(N+1) \supset H = \diag K \simeq SO(N+1)$.
The only change is that the isotropy subgroup of $H$ leaving $X$ fixed
is given by $U(1)\times O(N-1)$ 
which is a proper subgroup of $U(N-1)$. 
Nevertheless 
the Cartan structure equations of a $SO(N+1)$-parallel frame 
for non-stretching curve flows in $M\simeq SO(N+1)$
will actually turn out to exhibit a larger invariance
under unitary rotations $U(N-1)$.

\begin{remark}
We will set up parallel frames for curve flows 
in the Lie groups $G=SU(N),SO(N+1)$ 
using the same respective choice of unit vector $\xframe{}{}$
in $\vs{g}/\vs{so}(N) \subset \vs{g}=\vs{su}(N),\vs{so}(N+1)$
as was made in \cite{kievpaper} 
for curve flows in the corresponding symmetric spaces $G/SO(N)$. 
\end{remark}

\section{Bi-Hamiltonian operators and vector soliton equations for 
$\bdsymb{SU(N)}$}

Let $\gamma(t,x)$ be  a flow of a non-stretching curve in $G=SU(N)$. 
We consider a $SU(N)$-parallel coframe 
$\coframe{} \in T^*_\gamma G\otimes\vs{su}(N)$
and its associated connection $1$-form
$\conx{}{} \in T^*_\gamma G\otimes\vs{su}(N)$
along $\gamma$
\footnote{Note $\conx{}{}$ is related to $\coframe{}$ by
the Riemannian covariant derivative \eqref{ewrelation}
on the surface swept out by the curve flow $\gamma(t,x)$.}
given by
\EQ\label{suconx}
\xconx{}{} =\gamma_x \hook \conx{}{}
= \begin{pmatrix}
0 & \vvec\\ -\htrans{\vvec} & \bdsymb{0}
\end{pmatrix}
\in \vs{p}_{C^\perp},\qquad
\vvec \in \Cnum{N-1},\qquad
\bdsymb{0} \in \vs{u}(N-1)
\endEQ
and
\EQ\label{suframe}
\xframe{} = \gamma_x \hook \coframe{}
=\kappa\frac{\i}{N}
\begin{pmatrix}
1-N & \vec{0}\\ \trans{\vec{0}} & \bdsymb{1}
\end{pmatrix}
\in \vs{p}_\parallel  ,\qquad
\vec{0} \in \Rnum{N-1},\qquad
\i\bdsymb{1} \in \vs{u}(N-1)
\endEQ
up to a normalization factor $\kappa$ fixed as follows. 
Note the form of $\xframe{}$ indicates that 
the coframe is adapted to $\gamma$ provided 
$\xframe{}$ is scaled to have unit norm in the Cartan-Killing inner product, 
\EQ
<\xframe{},\xframe{}>_{\vs{p}}
= -\frac{1}{2} \tr\left(\kappa^2
\begin{pmatrix} N^{-1}-1 & 0\\ 0 & N^{-1}\bdsymb{1} \end{pmatrix}^2\right)
= \kappa^2 \frac{N-1}{2N}=1
\endEQ
by putting $\kappa^2=2 N(N-1)^{-1}$.
As a consequence, 
all $\vs{su}(N)$ matrices will have a canonical decomposition 
into tangential and normal parts relative to $\xframe{}$,
\EQs
\vs{su}(N) 
&=& 
\begin{pmatrix}
(N^{-1}-1) p_{\parallel}\i & \vec{p}_{\perp}\\ -\htrans{\vec{p}_{\perp}} &
\bdsymb{p_{\perp}}-N^{-1} p_{\parallel}\i\bdsymb{1}
\end{pmatrix}
\nonumber\\
&=& \frac{\i}{N}
\begin{pmatrix}
(1-N) p_{\parallel} & \vec{0}\\ \trans{\vec{0}} & p_{\parallel}\bdsymb{1}
\end{pmatrix}
+
\begin{pmatrix}
0 &\vec{p}_{\perp}\\ -\htrans{\vec{p}_{\perp}} &\bdsymb{p_{\perp}}
\end{pmatrix}
\simeq \vs{p}
\endEQs
parameterized by the matrix $\bdsymb{p_{\perp}} \in \vs{su}(N-1)$,
the vector $\vec{p}_{\perp} \in \Cnum{N-1}$,
and the scalar $\i p_{\parallel} \in \Rnum{}$,
corresponding to 
$\vs{p} = \vs{p}_{\parallel} \oplus \vs{p}_{\perp}$
where $\vs{p}_{\parallel} \simeq \vs{u}(1)$.
In this decomposition
the centralizer of $\xframe{}{}$ consists of matrices 
parameterized by $(p_\parallel,\bdsymb{p_\perp})$
and hence 
$\vs{p}_C \simeq \vs{u}(N-1) \supset \vs{p}_\parallel \simeq \vs{u}(1)$
while its perp space $\vs{p}_{C^\perp}\subset \vs{p}_\perp$
is parameterized by $\vec{p}_{\perp}$.
Note $\bdsymb{p_\perp}$ is empty only if $N=2$,
so consequently for $N>2$
the $SU(N)$-parallel frame \eqrefs{suconx}{suframe}
is a strict generalization of a Riemannian parallel frame.

In the flow direction we put
\EQs
\tframe{} = \gamma_{t} \hook \coframe{}
&=& \kappa \hpar\frac{\i}{N}
\begin{pmatrix}
1-N & \vec{0} \\ \trans{\vec{0}} & \bdsymb{1}
\end{pmatrix}
+\kappa
\begin{pmatrix}
0 &\hvec_{\perp} \\ -\htrans{\hvec_{\perp}} & \bdsymb{h}_{\perp}
\end{pmatrix} 
\in \vs{p}_\parallel \oplus \vs{p}_\perp 
\nonumber\\
&=&
\kappa
\begin{pmatrix}
(N^{-1}-1)\hpar\i & \hvec_{\perp} \\ -\htrans{\hvec_{\perp}} &
\bdsymb{h}_{\perp}+N^{-1}\hpar\i\bdsymb{1}
\end{pmatrix} 
\label{suet}
\endEQs
and
\EQs
\tconx{}{} = \gamma_{t} \hook \conx{}{}
=
\begin{pmatrix}
-\i\theta & \wvec \\ -\htrans{\wvec} & \bdsymb{\Theta}
\end{pmatrix}
\in \vs{p}_{C}\oplus\vs{p}_{C^\perp} ,
\label{suflowconx}
\endEQs
with
\EQs
&& 
\hpar\in\Rnum{},\qquad 
\hvec_{\perp} \in \Cnum{N-1} ,\qquad
\bdsymb{h}_{\perp} \in \vs{su}(N-1) ,
\nonumber\\
&&
\wvec \in \Cnum{N-1} ,\qquad
\bdsymb{\Theta} \in \vs{u}(N-1) ,\qquad
\theta=-\i\,\tr\bdsymb{\Theta} \in \Rnum{} .
\nonumber
\endEQs
The components $\hpar,(\hvec_{\perp},\bdsymb{h}_{\perp})$
correspond to decomposing
$\tframe{} =
g(\gamma_{t},\gamma_{x})\xframe{}+(\gamma_{t})_{\perp} \hook \coframe{\perp}$
into tangential and normal parts relative to $\xframe{}$.
We then have
\EQs
&& [\xframe{},\tframe{}] =
-\kappa^2 \i
\begin{pmatrix}
0 &\hvec_{\perp} \\ \htrans{\hvec_{\perp}} & \bdsymb{0}
\end{pmatrix}
\in \vs{p}_{C^\perp} ,
\\
&& [\xconx{}{},\tframe{}] =
\kappa
\begin{pmatrix}
\hvec_{\perp} \cdot \bar\vvec -\vvec \cdot \bar\hvec_{\perp} &
\vvec \hook \bdsymb{h}_{\perp} +\i\hpar \vvec \\
-\htrans{(\vvec \hook \bdsymb{h}_{\perp} + \i\hpar \vvec)} &
\bar\hvec_{\perp}\otimes \vvec -\bar\vvec \otimes \hvec_{\perp} 
\end{pmatrix}
\in \vs{p}_{C}\oplus\vs{p}_{C^\perp} ,
\\
&& [\tconx{}{},\xframe{}] =
\kappa \i
\begin{pmatrix}0 & \wvec \\ \htrans{\wvec} & \bdsymb{0}
\end{pmatrix}
\in \vs{p}_{C^\perp} .
\endEQs
Here $\otimes$ denotes the outer product of pairs of vectors
($1 \times N-1$ row matrices),
producing $N-1 \times N-1$ matrices
$\vec{A} \otimes \vec{B} = \trans{\vec{A}} \vec{B}$,
and $\hook$ denotes
multiplication of $N-1 \times N-1$ matrices
on vectors ($1 \times N-1$ row matrices),
$\vec{A} \hook (\vec{B} \otimes \vec{C}) := (\vec{A} \cdot \vec{B}) \vec{C}$
which is the transpose of the standard matrix product on column vectors,
$(\vec{B} \otimes \vec{C}) \vec{A} = (\vec{C} \cdot \vec{A}) \vec{B}$.

Hence the curvature equation \eqref{cartancurv} reduces to
\EQs
\D{t}\vvec - \D{x}\wvec - \vvec \hook\bdsymb{\Theta} -\i\theta\vvec 
&=&
-\kappa^2\i\hvec_{\perp} ,
\label{suveq}\\
\D{x}\bdsymb{\Theta}+\bar\wvec\otimes\vvec -\bar\vvec \otimes\wvec 
&=&
0 ,
\label{suthetaeq}\\
\i\D{x}\theta+\vvec\cdot\bar\wvec -\wvec\cdot\bar\vvec 
&=&
0 ,
\label{suthetatreq}
\endEQs
while the torsion equation \eqref{cartantors} yields
\EQs
(\frac{1}{N}-1)\i\D{x}\hpar 
+ \hvec_{\perp} \cdot\bar\vvec -\vvec\cdot\bar\hvec_{\perp} 
&=& 0 ,
\label{suhpareq}\\
\D{x}\bdsymb{h}_{\perp}
+\bar\hvec_{\perp}\otimes\vvec -\bar\vvec \otimes\hvec_{\perp}
-\frac{1}{N-1}( 
\vvec\cdot\bar\hvec_{\perp}-\hvec_{\perp}\cdot\bar\vvec )\bdsymb{1}
&=& 0 ,
\label{suheq}\\
\i\wvec - \D{x}\hvec_{\perp} -\i\hpar\vvec - \vvec \hook \bdsymb{h}_{\perp} ,
&=& 0 .
\label{suweq}
\endEQs

To proceed, we use equations
\eqref{suthetaeq}--\eqref{suheq} to eliminate
\EQs
\bdsymb{\Theta} &=&
\Dinv{x}( \bar\vvec\otimes \wvec -\bar\wvec\otimes \vvec ) ,
\label{sutheta}\\
\theta &=&
\i\Dinv{x}( \bar\wvec \cdot \vvec -\bar\vvec\cdot \wvec ) ,
\label{suthetatr}\\
\bdsymb{h}_{\perp} +\frac{1}{N} \hpar\i\bdsymb{1} &=&
\Dinv{x}( \bar\vvec\otimes\hvec_{\perp} -\bar\hvec_{\perp} \otimes\vvec ) ,
\\
(1-\frac{1}{N})\hpar &=&
\i\Dinv{x}(
\bar\hvec_{\perp}\cdot \vvec -\bar\vvec\cdot \hvec_{\perp} ) ,
\endEQs
in terms of the variables $\vvec$, $\hvec_{\perp}$, $\wvec$.
Then equation \eqref{suveq} gives a flow on $\vvec$,
\EQ
\vvec_t=
\D{x}\wvec 
+ \Dinv{x}(\wvec\cdot \bar\vvec-\vvec \cdot\bar\wvec)\vvec
+ \vvec\hook \Dinv{x}( \bar\vvec\otimes \wvec -\bar\wvec\otimes \vvec )
-\kappa^2\i\hvec_{\perp}
\endEQ
with
\EQ
\wvec =
-\i\D{x}\hvec_{\perp}
+\i\Dinv{x}( \bar\hvec_{\perp}\cdot\vvec -\bar\vvec \cdot\hvec_{\perp} )\vvec 
+ \i\vvec \hook \Dinv{x}( 
\bar\hvec_{\perp} \otimes\vvec -\bar\vvec\otimes \hvec_{\perp} )
\endEQ
obtained from equation \eqref{suweq}. 
We now read off the obvious operators
\EQ
\Hop =
\D{x}
-2\i\Dinv{x}(\bar\vvec{\cdot^\dagger \i}\ )\vvec
+\vvec \hook \Dinv{x}(\bar\vvec \wedge^\dagger\ ) ,\qquad
\Iop = \i ,
\label{suHop}
\endEQ
and introduce the related operator
\EQ
\Jop =\Iop^{-1}\circ\Hop\circ\Iop=
\D{x}+2\Dinv{x}(\bar\vvec \cdot^\dagger\ )\vvec 
+ \vvec \hook\Dinv{x}(\bar\vvec \odot^\dagger\ ) ,
\label{suJop}
\endEQ
where
$\vec{A} \wedge^\dagger \vec{B} :=
\vec{A} \otimes \vec{B} -\bar{\vec{B}} \otimes\bar{\vec{A}}$
is a Hermitian version of the wedge product 
$\vec{A} \wedge\vec{B} =
\vec{A} \otimes \vec{B} - \vec{B}\otimes\vec{A}$,
and where
$\vec{A} \odot^\dagger \vec{B} :=
\vec{A} \otimes \vec{B} +\bar{\vec{B}} \otimes\bar{\vec{A}}$
and 
$\vec{A} \cdot^\dagger \vec{B} := 
\frac{1}{2}\vec{A} \cdot\vec{B}
+\frac{1}{2}\bar{\vec{B}}\cdot\bar{\vec{A}}$
are Hermitian versions of the symmetric product 
$\vec{A} \odot\vec{B} = \vec{A} \otimes\vec{B}+\vec{B}\otimes\vec{A}$
and dot product
$\vec{A} \cdot\vec{B}$. 
Note the intertwining
$\vec{A}\odot^\dagger \i\vec{B} = \i(\vec{A}\wedge^\dagger\vec{B})$ 
and vice versa
$\vec{A}\wedge^\dagger \i\vec{B} = \i(\vec{A}\odot^\dagger\vec{B})$,
which imply the identities
$\bar{\vec{A}}\wedge^\dagger \vec{A} 
= \bar{\vec{A}}\odot^\dagger \i\vec{A} =0$,
$\bar{\vec{A}} \cdot^\dagger \i\vec{A} = 0$,
and 
$\tr(\bar{\vec{A}} \odot^\dagger \vec{A}) = 
-\i\,\tr(\bar{\vec{A}}\wedge^\dagger \i\vec{A}) 
= 2\vec{A}\cdot\bar{\vec{A}} =2|\vec{A}|^2$.

The Hamiltonian structure determined by these operators $\Hop,\Jop,\Iop$
and variables $\vvec,\wvec,\hvec_\perp$
in the space $G=SU(N)$ is somewhat different 
in comparison to the space $G/SO(N)$. 
Use of the methods in \cite{SandersWang2} establishes 
the following main result. 

\begin{theorem}\label{thm1}
$\Hop, \Iop$ are a Hamiltonian pair of 
$U(N-1)$-invariant cosymplectic operators with respect to
the Hamiltonian variable $\vvec$,
while $\Jop,\Iop^{-1}=-\Iop$ 
are compatible symplectic operators. 
Consequently, the flow equation
takes the Hamiltonian form
\EQ
\vvec_{t} = \Hop(\wvec) - \chi \Iop(\hvec_{\perp}) =
\Rop^2(\i\hvec_{\perp}) - \chi \i\hvec_{\perp} ,\qquad
-\wvec = \Jop(\i\hvec_{\perp})=\i\Hop(\hvec_{\perp})
\endEQ
where $\Rop = \Hop\circ\Iop = \Iop\circ\Jop$ 
is a hereditary recursion operator.
\end{theorem}

Here $\chi=\kappa^2$ is a constant related to 
the Riemannian scalar curvature of the space $G=SU(N)$. 

On the $x$-jet space of $\vvec(t,x)$, the variables $\i\hvec_{\perp}$
and $\wvec$ have the respective meaning of a Hamiltonian vector field
$\i\hvec_{\perp}\hook \partial/\partial \vvec$
and covector field $\wvec\hook d\vvec$
(see \cite{Dorfman,Olver} and the appendix of \cite{JPhysApaper}).
Thus the recursion operator
\EQ
\Rop=
\i( \D{x} + 2\Dinv{x}(\bar\vvec \cdot^\dagger\ ) \vvec
+ \vvec \hook \Dinv{x}(\bar\vvec \odot^\dagger\ ) )
\label{suRop}
\endEQ
generates a hierarchy of commuting Hamiltonian vector fields
$\i\hvec^{(k)}_{\perp}$ starting from $\i\hvec^{(0)}_{\perp}=\i\vvec$
defined by the infinitesimal generator of phase rotations on $\vvec$,
and followed by $\i\hvec^{(1)}_{\perp}=-\vvec_x$ 
which is the infinitesimal generator of $x$-translations on $\vvec$
(in terms of arclength $x$ along the curve). 

The adjoint operator 
\EQ
\Rop^*=
\i\D{x} + 2\Dinv{x}(\bar\vvec {\cdot^\dagger\i}\ ) \vvec
+ \i\vvec \hook \Dinv{x}(\bar\vvec \wedge^\dagger\ ) )
\endEQ
generates a related hierarchy of
involutive Hamiltonian covector fields
$\wvec^{(k)} = \delta H^{(k)}/\delta \bar\vvec$
in terms of Hamiltonians
$H =H^{(k)}(\vvec,\vvec_{x},\vvec_{2x},\ldots)$ 
starting from 
$\wvec^{(0)}=-\vvec$, 
$H^{(0)}=-\vvec\cdot\bar\vvec=-|\vvec|^2$,
and followed by 
$\wvec^{(1)}=-\i\vvec_x$, 
$H^{(1)}=\frac{1}{2}\i(\vvec\cdot\bar\vvec_x-\vvec_x\cdot\bar\vvec)
= \vvec\cdot^\dagger \i\bar\vvec_x$. 
These hierarchies are
related by 
\EQ\label{hwhierarchy}
\i\hvec^{(k+1)}_{\perp} = \Hop(\wvec^{(k)}), \qquad
\wvec^{(k+1)}=-\Jop(\i\hvec^{(k)}_{\perp}), \qquad
\hvec^{(k)}_{\perp} = -\wvec^{(k)}, 
\endEQ
for $k=0,1,2,\ldots$,
so thus $\i\wvec^{(k)}$ can be also interpreted as a Hamiltonian vector field
and $\hvec^{(k)}_{\perp}$ as a corresponding covector field. 
Both hierarchies share the NLS scaling symmetry
$x\rightarrow\lambda x$, $\vvec\rightarrow\lambda^{-1}\vvec$,
under which we see
$\hvec^{(k)}_{\perp}$ and $\wvec^{(k)}$ have scaling weight $1+k$,
while $H^{(k)}$ has scaling weight $2+k$.

\begin{corollary}\label{cor1}
Associated to the recursion operator $\Rop$ there is a
corresponding hierarchy of commuting bi-Hamiltonian flows on $\vvec$
given by $U(N-1)$-invariant vector evolution equations
\EQ\label{floweq}
\vvec_{t} = \i(\hvec^{(k+2)}_{\perp} - \chi \hvec^{(k)}_{\perp})
=\Hop(\delta H^{(k+1,\chi)}/\delta \bar\vvec)
=-\Iop(\delta H^{(k+2,\chi)}/\delta \bar\vvec) ,\qquad
\endEQ
with Hamiltonians $H^{(k+2,\chi)} = H^{(k+2)} -\chi H^{(k)}$,
$k = 0,1,2,\ldots$.
In the terminology of \cite{kievpaper},
$\hvec^{(k)}_{\perp}$ will be said to produce a $+(k+1)$ flow on $\vvec$.
\end{corollary}

The $+1$ flow given by $\hvec_{\perp}=\vvec$ yields
\EQ\label{sunlseq}
\i\vvec_t =
\vvec_{2x}+2|\vvec|^2\vvec +\chi\vvec
\endEQ
which is a vector NLS equation up to a phase term 
that can be absorbed by a phase rotation 
$\vvec \rightarrow \exp(-\i\chi t)\vvec$.
Higher-order versions of this equation are produced by the
$+(1+2k)$ odd-flows, $k\ge 1$. 

The $+2$ flow is given by $\hvec_{\perp}=\i\vvec_x$, 
yielding a complex vector mKdV equation 
\EQ\label{sumkdveq}
\vvec_t =
\vvec_{3x}+3|\vvec|^2\vvec_x+3(\vvec_x\cdot\bar\vvec)\vvec + \chi\vvec_x 
\endEQ
up to a convective term that can be absorbed by 
a Galilean transformation $x \rightarrow x +\chi t$, $t\rightarrow t$.
The $+(2+2k)$ even-flows, $k\ge 1$, 
give higher-order versions of this equation. 

There is also a $0$ flow $\vvec_{t} = \vvec_{x}$ arising from
$\hvec_{\perp}=0$, $\hpar=1$, which falls outside the hierarchy
generated by $\Rop$.

All these flows correspond to geometrical motions of the curve $\gamma(t,x)$,
given by
\EQ
\gamma_{t}=
f(\gamma_{x},\covder{x}\gamma_{x},\covder{x}^{2}\gamma_{x},\ldots)
\endEQ
subject to the non-stretching condition
\EQ
|\gamma_{x}|_{g}=1 .
\endEQ
The equation of motion for $\gamma$  is obtained
from the identifications
$\gamma_{t} \leftrightarrow \tframe{}$,
$\covder{x}\gamma_{x} \leftrightarrow \covD{x}\xframe{}
= [\xconx{}{},\xframe{}]$,
and so on, using
$\covder{x} \leftrightarrow \D{x} + [\xconx{}{},\cdot] = \covD{x}$.
These identifications correspond to 
$T_x G \leftrightarrow \vs{su}(N)\simeq \vs{p}$
as defined via the parallel coframe
along $\gamma$ in $G=SU(N)$. 
Note we have
\EQs
[\xconx{}{},\xframe{}] &=&
\kappa\i
\begin{pmatrix}
0 & \vvec \\\htrans{\vvec} & \bdsymb{0}
\end{pmatrix} ,
\\ \nonumber
[\xconx{}{},[\xconx{}{},\xframe{}]] &=&
2\kappa\i
\begin{pmatrix}
|\vvec|^2 & \vec{0} \\ \vec{0} & -\bar\vvec \otimes \vvec
\end{pmatrix} ,
\endEQs
and so on.
In addition,
\EQs
\ad([\xconx{}{},\xframe{}])\xframe{} &=&
-\kappa^2
\begin{pmatrix}
0 & \vvec \\ -\htrans{\vvec} & \bdsymb{0}
\end{pmatrix} 
= -\ad(\xframe{})[\xconx{}{},\xframe{}] ,
\\ 
\ad([\xconx{}{},\xframe{}])^2\xframe{} &=&
2\kappa^3\i
\begin{pmatrix}|
\vvec|^2 & \vec{0} \\ \vec{0} & -\bar\vvec \otimes \vvec
\end{pmatrix}
\\ \nonumber
&=& \chi [\xconx{}{},[\xconx{}{},\xframe{}]] ,
\nonumber
\endEQs
where $\ad(\cdot)$ denotes the standard adjoint representation acting in the
Lie algebra $\vs{su}(N)$.

For the $+1$ flow,
\EQ
\hvec_{\perp}=\vvec ,\qquad
\hpar= 0 ,\qquad
\bdsymb{h}_{\perp}= \bdsymb{0} ,
\endEQ
we have (from equation \eqref{suet})
\EQs
\tframe{} &=&
\kappa
\begin{pmatrix}
(N^{-1}-1)\hpar\i & \hvec_{\perp} \\ -\htrans{\hvec_{\perp}} &
\bdsymb{h}_{\perp}+N^{-1}\hpar\i\bdsymb{1}
\end{pmatrix}
=\kappa
\begin{pmatrix}
0 & \vvec \\ -\htrans{\vvec} & \bdsymb{0}
\end{pmatrix}
\nonumber \\ 
&=&
\chi^{-1/2}\ad(\xframe{})[\xconx{}{},\xframe{}] 
\label{sunlsframeflow}
\endEQs
which we identify as $J_\gamma\covder{x}\gamma_x$
where $J_\gamma \leftrightarrow \ad(\xframe{})$
is an algebraic operator in $T_x G \leftrightarrow \vs{su}(N)$
obeying $J_\gamma^2 =-{\rm id}$.
Hence the frame equation \eqref{sunlsframeflow} 
describes a geometric nonlinear PDE for $\gamma(t,x)$
\EQ\label{nlsmap}
\gamma_t = \chi^{-1/2} J_\gamma\covder{x}\gamma_x ,\qquad 
J_\gamma=\ad(\gamma_x)
\endEQ
analogous to the well-known Schr\"odinger map. 
This PDE \eqref{nlsmap}
will be called a {\it chiral Schr\"odinger map} equation 
on the Lie group $G=SU(N)$.
(A different derivation was given in \cite{JPhysApaper}
using left-invariant frames, 
which needed a Lie algebraic restriction on the curve $\gamma$. 
Here we see any such restrictions are unnecessary, 
due to the use of a parallel frame.)

Next, for the $+2$ flow
\EQ
\hvec_{\perp}=\i\vvec_x ,\qquad
\hpar=
N(N-1)^{-1} |\vvec|^2 ,\qquad
\bdsymb{h}_{\perp}=
\i( \bar\vvec \otimes \vvec +(1-N)^{-1} |\vvec|^2 \bdsymb{1} ) ,
\endEQ
we obtain (again via equation \eqref{suet})
\EQs
\tframe{} &=&
\kappa
\begin{pmatrix}
(N^{-1}-1)\hpar\i & \hvec_{\perp} \\ -\htrans{\hvec_{\perp}} &
\bdsymb{h}_{\perp}+N^{-1}\hpar\i\bdsymb{1}
\end{pmatrix}
=\kappa\i
\begin{pmatrix}
-|\vvec|^2 & \vvec_x \\\htrans{\vvec_x} & \bar\vvec \otimes \vvec
\end{pmatrix}
\\ \nonumber &=&
\D{x}[\xconx{}{},\xframe{}]
-\frac{1}{2}[\xconx{}{},[\xconx{}{},\xframe{}]] .
\nonumber
\endEQs
Then writing these expressions in terms of $\covD{x}$ and
$\ad([\xconx{}{},\xframe{}])$, we get
\EQ
\tframe{} =
\covD{x}[\xconx{}{},\xframe{}]
-\frac{3}{2}\chi^{-1} \ad([\xconx{}{},\xframe{}])^2\xframe{} .
\endEQ
The first term is identified as
$\covD{x}[\xconx{}{},\xframe{}] 
\leftrightarrow
\covder{x}(\covder{x}\gamma_x)=\covder{x}^2\gamma_x$.
For the second term we observe
$\ad([\xconx{}{},\xframe{}])^2 
\leftrightarrow
\chi^{-1}\ad(\covder{x}\gamma_x)^2$.
Thus, 
$\gamma(t,x)$ satisfies a geometric nonlinear PDE 
\EQ\label{mkdvmap}
\gamma_t =
\covder{x}^2 \gamma_x
- \frac{3}{2}\chi^{-1}\ad(\covder{x} \gamma_x)^2 \gamma_x 
\endEQ
called the non-stretching mKdV map equation 
on the Lie group $G=SU(N)$.
We will refer to it as the {\it chiral mKdV map}. 
The same PDE was found to arise from curve flows in 
the corresponding symmetric space $G/SO(N)$. 

All higher odd- and even- flows on $\vvec$ in the hierarchy 
respectively determine higher-order 
chiral Schr\"odinger map equations and chiral mKdV map equations for $\gamma$.
The $0$ flow on $\vvec$ directly corresponds to
\EQ\label{convmap}
\gamma_t=\gamma_x
\endEQ
which is just a convective (linear traveling wave) map equation.

In addition there is a $-1$ flow contained in the hierarchy,
with the property that $\hvec_{\perp}$ is annihilated by
the symplectic operator $\Jop$
and hence gets mapped into $\Rop(\hvec_{\perp})=0$
under the recursion operator.
Geometrically this flow means simply
$\Jop(\hvec_{\perp})=\wvec=0$
which implies $\tconx{}{}=0$
from equations \eqref{suflowconx}, \eqref{sutheta}, \eqref{suthetatr},
and hence
$0=[\tconx{}{},\xframe{}]=\covD{t}\xframe{}$
where
$\covD{t} = \D{t}+ [\tconx{}{},\cdot]$.
The correspondence 
$\covder{t} \leftrightarrow \covD{t}$, 
$\gamma_x \leftrightarrow \xframe{}$ 
immediately leads to the equation of motion
\EQ\label{wavemap}
0 = \covder{t}\gamma_x
\endEQ
for the curve $\gamma(t,x)$.
This nonlinear geometric PDE is recognized to be 
a non-stretching wave map equation on the Lie group $G=SU(N)$,
which also was found to arise \cite{kievpaper} in the same manner 
from curve flows in $G/SO(N)$. 

The $-1$ flow equation produced on $\vvec$ is a nonlocal evolution equation
\EQ\label{suvsgflow}
\vvec_t=-\chi\i \hvec_{\perp} ,\qquad
\chi=\kappa^2
\endEQ
with $\hvec_{\perp}$ satisfying
\EQ\label{suwsgflow}
0 = \i\wvec =\D{x} \hvec_{\perp} + \i h\vvec +\vvec \hook\bdsymb{h} ,
\endEQ
where it is convenient to introduce the variables
\EQ
\bdsymb{h} = \bdsymb{h}_{\perp}+N^{-1}\hpar\i\bdsymb{1}, \qquad
h=N^{-1}(N-1)\hpar=-\i\,\tr\bdsymb{h} ,
\endEQ
which satisfy
\EQs
\D{x}h &=&
\i( \bar\hvec_{\perp} \cdot\vvec-\bar\vvec \cdot\hvec_{\perp} ) ,
\label{suhparsgflow}\\
\D{x} \bdsymb{h} &=&
\bar\vvec \otimes\hvec_{\perp} -\bar\hvec_{\perp} \otimes \vvec .
\label{suhsgflow}
\endEQs
Note these variables $\hvec_{\perp}$, $h$, $\bdsymb{h}$
will be nonlocal functions of $\vvec$ (and its $x$ derivatives)
as determined by equations \eqsref{suwsgflow}{suhsgflow}. 
To proceed, as in the case $G/SO(N)$, 
we seek an inverse local expression for $\vvec$ 
arising from an algebraic reduction of the form
\EQ\label{suhsgeq}
\bdsymb{h} 
= \alpha\i\bar\hvec_{\perp} \otimes\hvec_{\perp} +\beta\i \bdsymb{1} 
\endEQ
for some expressions $\alpha(h),\beta(h)\in\Rnum{}$. 
Similarly to the analysis for the case $G/SO(N)$, 
substitution of $\bdsymb{h}$ into equation \eqref{suhsgflow} 
followed by the use of equations \eqrefs{suwsgflow}{suhparsgflow}
yields
\EQ\label{sualphaeq}
\alpha = -(h+\beta)^{-1} ,\qquad 
\beta =\const.
\endEQ
Next, by taking the trace of $\bdsymb{h}$ from equation \eqref{suhsgeq},
we obtain
\EQ\label{suhperpsgeq}
|\hvec_{\perp}|^2 = N\beta (h+\beta)-(h+\beta)^2
\endEQ
which enables $h$ to be expressed in terms of
$\hvec_{\perp}$ and $\beta$.
To determine $\beta$
we use the conservation law 
\EQ
0 =\D{x}( |\hvec_{\perp}|^2 +\frac{1}{2}( h^2+|\bdsymb{h}|^2) ) ,
\endEQ
admitted by equations \eqsref{suwsgflow}{suhsgflow},
corresponding to a wave map conservation law 
\EQ\label{wavemapconslaw}
0=\covder{x}|\gamma_t|_g^2
\endEQ
where
\EQ
|\gamma_t|_{g}^{2}
= <\tframe{},\tframe{}>_\vs{p}
= \kappa^2( |\hvec_{\perp}|^2 +\frac{1}{2}( h^2+|\bdsymb{h}|^2) )
\endEQ
and 
\EQ
|\bdsymb{h}|^2 := -\tr( \bdsymb{h}^2 )
= \alpha^2 |\hvec_{\perp}|^4 +2\alpha \beta |\hvec_{\perp}|^2 +\beta^2 (N-1) .
\endEQ
A conformal scaling of $t$ can now be used to make
$|\gamma_t|_{g}$ equal to a constant.
To simplify subsequent expressions we put 
$|\gamma_t|_{g}=2$, so then
\EQ
(2/\kappa)^2
=|\hvec_{\perp}|^2 +\frac{1}{2}( |\bdsymb{h}|^2+h^2 ) .
\endEQ
Substitution of equations \eqsref{suhsgeq}{suhperpsgeq}
into this expression yields
\EQ
\beta^2 = (2/N)^2
\endEQ
from which we obtain via equation \eqref{suhperpsgeq}
\EQ\label{suhparalpha}
h= 2N^{-1} -1 \pm \sqrt{1-|\hvec_{\perp}|^2} ,\qquad
\alpha =|\hvec_{\perp}|^{-2}
( 1 \pm \sqrt{1-|\hvec_{\perp}|^2} ) .
\endEQ
These variables then can be expressed in terms of $\vvec$
through the flow equation \eqref{suvsgflow},
\EQ
|\hvec_{\perp}|^2 =\chi^{-2} |\vvec_t|^2 .
\endEQ
In addition, 
by substitution of equations \eqrefs{suhsgeq}{sualphaeq}
into equation \eqref{suwsgflow}
combined with the relation \eqref{suvsgflow}, we obtain 
\EQ
\hvec_{\perp} = 
\i\Dinv{x}( \alpha^{-1} \vvec 
-\chi^{-2}\alpha (\vvec\cdot\bar\vvec_t) \vvec_t ) .
\endEQ
Finally, the same equations also yield the inverse expression
\EQ\label{suvhsgeq}
\i\vvec = 
\alpha( \hvec_\perp{}_x 
+\i\alpha(\bar\hvec_\perp\cdot\vvec) \hvec_\perp ) ,\quad
\i\bar\hvec_\perp\cdot\vvec
= \alpha (1-\alpha^2 \bar\hvec_{\perp}\cdot\hvec_{\perp})^{-1}
\bar\hvec_{\perp} \cdot\hvec_{\perp}{}_x 
\endEQ
after a dot product is taken with $\bar\hvec_{\perp}$. 

Hence, with the factor $\chi$ absorbed by a scaling of $t$,
the $-1$ flow equation on $\vvec$ becomes
the nonlocal evolution equation 
\EQ
\vvec_t = 
\Dinv{x}( 
A_{\mp}\vvec -A_{\pm} |\vvec_t|^{-2} (\vvec\cdot\bar\vvec_t) \vvec_t )
\endEQ
where
\EQ
A_{\pm} := 1 \pm \sqrt{1-|\vvec_t|^2} =|\vvec_t|^2 /A_{\mp} .
\endEQ
In hyperbolic form
\EQ\label{susghyperboliceq}
\vvec_{tx}=
A_{\pm}\vvec - A_{\mp}|\vvec_t|^{-2}(\vvec \cdot \bar\vvec_t)\vvec_t 
\endEQ
gives a complex variant of a vector SG equation,
found in \cite{AncoWolf}. 
Equivalently, through relations \eqrefs{suvhsgeq}{suhparalpha},
$\hvec_{\perp}$ is found to obey a complex vector SG equation
\EQ\label{suSGeq}
( \alpha( \hvec_{\perp}{}_x 
\mp \frac{1}{2}\alpha (1- |\hvec_{\perp}|^2)^{-1/2} 
(\bar\hvec_{\perp}\cdot \hvec_{\perp}{}_x) \hvec_{\perp} ) )_t
=\hvec_{\perp} .
\endEQ

It is known from the symmetry-integrability classification results
in \cite{AncoWolf} 
that the hyperbolic vector equation \eqref{susghyperboliceq}
admits the vector NLS equation \eqref{sunlseq} as a higher symmetry.
As a consequence of Corollary~\ref{cor1}, 
the hierarchy of vector NLS/mKdV higher symmetries
\EQs
\vvec_{t}^{(0)} &=&
\i\vvec ,
\label{su0flow}\\
\vvec_{t}^{(1)} &=& \Rop(\i\vvec) 
= -\vvec_x ,
\label{su1flow}\\
\vvec_{t}^{(2)} &=& \Rop^2(\i\vvec) =
-\i( \vvec_{2x}+2|\vvec|^2\vvec ) ,
\label{su2flow}\\
\vvec_{t}^{(3)} &=& \Rop^2(-\vvec_x) =
\vvec_{3x}+3|\vvec|^2\vvec_x+3(\vvec_x\cdot\bar\vvec)\vvec ,
\label{su3flow}
\endEQs
and so on,
generated by the recursion operator \eqref{suRop},
all commute with 
the $-1$ flow
\EQ\label{su-1flow}
\vvec_{t}^{(-1)} = -\i\hvec_{\perp}
\endEQ
associated to the vector SG equation \eqref{suSGeq}.
Therefore all these symmetries are admitted by 
the hyperbolic vector equation \eqref{susghyperboliceq}. 
The corresponding hierarchy of NLS/mKdV Hamiltonians
(modulo total derivatives)
\EQs
H^{(0)} &=&
|\vvec|^2 ,
\nonumber\\
H^{(1)} &=&
\i\vvec_x\cdot\bar\vvec ,
\nonumber\\
H^{(2)} &=&
-|\vvec_x|^2 +|\vvec|^4 ,
\nonumber\\
H^{(3)} &=&
\i\vvec_x\cdot(\bar\vvec_{2x}+3|\vvec|^2 \bar\vvec) ,
\nonumber
\endEQs
and so on,
generated from the adjoint recursion operator, 
are all conserved densities for the $-1$ flow and
thereby determine conservation laws admitted 
for the hyperbolic vector equation \eqref{susghyperboliceq}.

Viewed as flows on $\vvec$, 
the entire hierarchy of vector PDEs
\eqsref{su0flow}{su3flow}, etc.,
including the $-1$ flow \eqref{su-1flow}, 
possesses the NLS scaling symmetry
$x\rightarrow\lambda x$, $\vvec\rightarrow\lambda^{-1}\vvec$,
with $t\rightarrow\lambda^{k} t$
for $k=-1,0,1,2$, etc..
As well, 
the flows for $k\geq 0$ will be local polynomials in the variables
$\vvec,\vvec_x,\vvec_{2x},\ldots$
as established by general results in \cite{Wang-thesis,Sergyeyev,Sergyeyev2}
concerning nonlocal operators. 

\begin{theorem}\label{thm3}
In the Lie group $SU(N)$
there is a hierarchy of bi-Hamiltonian flows of curves $\gamma(t,x)$
described by geometric map equations.
The $0$ flow is a convective (traveling wave) map \eqref{convmap},
while the $+1$ flow is a non-stretching 
chiral Schr\"odinger map \eqref{nlsmap}
and the $+2$ flow is a non-stretching 
chiral mKdV map \eqref{mkdvmap},
and the other odd- and even- flows are higher order analogs.
The kernel of the recursion operator \eqref{suRop} in the hierarchy
yields the $-1$ flow which is a non-stretching chiral wave map \eqref{wavemap}.
Moreover the components of the principal normal vector along the
$+1,+2,-1$ flows in a $SU(N)$-parallel frame respectively satisfy 
a vector NLS equation \eqref{sunlseq}, 
a complex vector mKdV equation \eqref{sumkdveq}
and a complex vector hyperbolic equation \eqref{susghyperboliceq}. 
\end{theorem}

\section{Bi-Hamiltonian operators and vector soliton equations for 
$\bdsymb{SO(N+1)}$}

Let $\gamma(t,x)$ be  a flow of a non-stretching curve in $G=SO(N+1)$. 
We introduce a $SO(N+1)$-parallel coframe 
$\coframe{} \in T^*_\gamma G\otimes\vs{so}(N+1)$
and its associated connection $1$-form
$\conx{}{} \in T^*_\gamma G\otimes\vs{so}(N+1)$
along $\gamma$
\footnote{As before, $\conx{}{}$ is related to $\coframe{}$ by
the Riemannian covariant derivative \eqref{ewrelation}
on the surface swept out by the curve flow $\gamma(t,x)$.}
\EQ\label{soconx}
\xconx{}{} =\gamma_x \hook \conx{}{}
= \begin{pmatrix}
0 & 0 & \vvec_1 \\ 0 & 0 & \vvec_2\\ 
-\trans{\vvec_1} & -\trans{\vvec_2} & \bdsymb{0}
\end{pmatrix}
\in \vs{p}_{C^\perp},\qquad
\vvec_1, \vvec_2 \in \Rnum{N-1}
\endEQ
and
\EQ\label{soframe}
\xframe{} = \gamma_x \hook \coframe{}
=\begin{pmatrix}
0 & 1 & \vec{0}\\ -1 & 0 & \vec{0}\\ 
\trans{\vec{0}} & \trans{\vec{0}} & \bdsymb{0}
\end{pmatrix}
\in \vs{p}_\parallel  ,\qquad
\vec{0} \in \Rnum{N-1},\qquad
\bdsymb{0} \in \vs{so}(N-1)
\endEQ
normalized so that 
$\xframe{}$ has unit norm in the Cartan-Killing inner product, 
$
<\xframe{},\xframe{}>_{\vs{p}}
= -\frac{1}{2} \tr\Big(
{\begin{pmatrix} 0 & 1\\ -1 & 0\end{pmatrix}}^2 \Big)
= 1$
indicating that the coframe is adapted to $\gamma$. 
Consequently, 
all $\vs{so}(N+1)$ matrices will have a canonical decomposition 
into tangential and normal parts relative to $\xframe{}$,
\EQs
\vs{so}(N+1) 
&=& 
\begin{pmatrix}
0 & p_\parallel & \vec{p}_{1\perp} \\ 
-p_\parallel & 0 &\vec{p}_{2\perp} \\ 
-\trans{\vec{p}_{1\perp}} & -\trans{\vec{p}_{2\perp}} & \bdsymb{p_\perp}
\end{pmatrix}
\simeq \vs{p}
\endEQs
parameterized by the matrix $\bdsymb{p_{\perp}} \in \vs{so}(N-1)$,
the pair of vectors $\vec{p}_{1\perp},\vec{p}_{2\perp} \in \Rnum{N-1}$,
and the scalar $p_{\parallel} \in \Rnum{}$,
corresponding to 
$\vs{p} = \vs{p}_{\parallel} \oplus \vs{p}_{\perp}$
where $\vs{p}_{\parallel} \simeq\vs{so}(2) \simeq \vs{u}(1)$.
The centralizer of $\xframe{}{}$ thus consists of matrices 
parameterized by $(p_\parallel,\bdsymb{p_\perp})$
and hence 
$\vs{p}_C \simeq \vs{u}(1)\oplus\vs{so}(N-1) 
\supset \vs{p}_\parallel \simeq \vs{u}(1)$
while its perp space $\vs{p}_{C^\perp}\subset \vs{p}_\perp$
is parameterized by $\vec{p}_{1\perp},\vec{p}_{2\perp}$.
As before, $\bdsymb{p_\perp}$ is empty only if $N=2$,
so consequently for $N>2$
the $SO(N+1)$-parallel frame \eqrefs{soconx}{soframe}
is a strict generalization of a Riemannian parallel frame.

In the flow direction we put
\EQs
\tframe{} = \gamma_{t} \hook \coframe{}
&=& 
\hpar
\begin{pmatrix}
0 & 1 & \vec{0}\\ -1 & 0 & \vec{0}\\ 
\trans{\vec{0}} & \trans{\vec{0}} & \bdsymb{0}
\end{pmatrix}
+ 
\begin{pmatrix}
0 & 0 & \hvec_{1\perp} \\ 
0 & 0 &\hvec_{2\perp} \\ 
-\trans{\hvec_{1\perp}} & -\trans{\hvec_{2\perp}} & \bdsymb{h_\perp}
\end{pmatrix}
\in \vs{p}_\parallel \oplus \vs{p}_\perp 
\nonumber\\
&=&
\begin{pmatrix}
0 & \hpar & \hvec_{1\perp} \\ 
-\hpar & 0 &\hvec_{2\perp} \\ 
-\trans{\hvec_{1\perp}} & -\trans{\hvec_{2\perp}} & \bdsymb{h_\perp}
\end{pmatrix} 
\label{soet}
\endEQs
and
\EQs
\tconx{}{} = \gamma_{t} \hook \conx{}{}
=
\begin{pmatrix}
0 & \theta & \wvec_1 \\ 
-\theta & 0 &\wvec_2 \\ 
-\trans{\wvec_1} & -\trans{\wvec_2} & \bdsymb{\Theta}
\end{pmatrix}
\in \vs{p}_{C}\oplus\vs{p}_{C^\perp} , 
\label{soflowconx}
\endEQs
with
\EQs
&& 
\hpar\in\Rnum{},\qquad 
\hvec_{1\perp},\hvec_{2\perp}, \in \Rnum{N-1} ,\qquad
\bdsymb{h}_{\perp} \in \vs{so}(N-1) ,
\nonumber\\
&&
\wvec_1,\wvec_2 \in \Rnum{N-1} ,\qquad
\bdsymb{\Theta} \in \vs{so}(N-1) ,\qquad
\theta \in \Rnum{} .
\nonumber
\endEQs
Here the components $\hpar,(\hvec_{1\perp},\hvec_{2\perp},\bdsymb{h}_{\perp})$
correspond to decomposing
$\tframe{} =
g(\gamma_{t},\gamma_{x})\xframe{}+(\gamma_{t})_{\perp} \hook \coframe{\perp}$
into tangential and normal parts relative to $\xframe{}$.
We now have, in the same notation used before, 
\EQs
&& [\xframe{},\tframe{}] 
=
\begin{pmatrix}
0 & 0 & \hvec_{2\perp} \\ 
0 & 0 & -\hvec_{1\perp} \\ 
-\trans{\hvec_{2\perp}} & \trans{\hvec_{1\perp}} & \bdsymb{0}
\end{pmatrix}
\in \vs{p}_{C^\perp} ,
\\
&& [\xconx{}{},\tframe{}] 
=
\nonumber\\&&
\begin{pmatrix}
0 & \vvec_2\cdot\hvec_{1\perp} -\vvec_1\cdot\hvec_{2\perp} 
& \vvec_1\hook\bdsymb{h_\perp} -\hpar\vvec_2 \\ 
\vvec_1\cdot\hvec_{2\perp} -\vvec_2\cdot\hvec_{1\perp} & 0 
& \vvec_2\hook\bdsymb{h_\perp} +\hpar\vvec_1 \\ 
-\trans{(\vvec_1\hook\bdsymb{h_\perp})} + \hpar\trans{\vvec_2} 
& -\trans{(\vvec_2\hook\bdsymb{h_\perp})} -\hpar\trans{\vvec_1} 
&\begin{matrix}
\hvec_{1\perp}\otimes\vvec_1 + \hvec_{2\perp}\otimes\vvec_2 \\
- \vvec_1 \otimes\hvec_{1\perp} - \vvec_2 \otimes\hvec_{2\perp}\end{matrix}
\end{pmatrix}
\nonumber\\ 
&&\qquad
\in \vs{p}_{C}\oplus\vs{p}_{C^\perp} ,
\\
&& [\tconx{}{},\xframe{}] 
= 
\begin{pmatrix}
0 & 0 & -\wvec_2 \\ 
0 & 0 & \wvec_1 \\ 
\trans{\wvec_2} & -\trans{\wvec_1} & \bdsymb{0}
\end{pmatrix}
\in \vs{p}_{C^\perp} .
\endEQs

The resulting torsion and curvature equations can be simplified
if we adopt a complex variable notation
\EQ
\vvec:= \vvec_1 +\i\vvec_2 ,\qquad
\wvec:= \wvec_1 +\i\wvec_2 ,\qquad
\hvec_\perp:= \hvec_{1\perp} +\i\hvec_{2\perp} .
\endEQ
Hence the curvature equation \eqref{cartancurv} becomes
\EQs
\D{t}\vvec - \D{x}\wvec - \vvec \hook\bdsymb{\Theta} -\i\theta\vvec 
&=&
-\i\hvec_{\perp} ,
\label{soveq}\\
2\D{x}\bdsymb{\Theta}+\bar\wvec\otimes\vvec +\wvec\otimes\bar\vvec 
-\bar\vvec \otimes\wvec -\vvec \otimes\bar\wvec 
&=&
0 ,
\label{sothetaeq}\\
2\i\D{x}\theta+\vvec\cdot\bar\wvec-\wvec\cdot\bar\vvec &=&
0 ,
\label{sothetatreq}
\endEQs
and the torsion equation \eqref{cartantors} reduces to
\EQs
2\i\D{x}\hpar 
-\hvec_{\perp} \cdot\bar\vvec +\vvec\cdot\bar\hvec_{\perp} 
&=& 0 ,
\label{sohpareq}\\
2\D{x}\bdsymb{h}_{\perp}
+\bar\hvec_{\perp}\otimes\vvec 
+\hvec_{\perp}\otimes\bar\vvec 
-\bar\vvec \otimes\hvec_{\perp}
-\vvec \otimes\bar\hvec_{\perp}
&=& 0 ,
\label{soheq}\\
\i\wvec - \D{x}\hvec_{\perp} -\i\hpar\vvec - \vvec \hook \bdsymb{h}_{\perp} ,
&=& 0 .
\label{soweq}
\endEQs
These equations are nearly the same as those for the space $G=SU(N)$,
except that both $\bdsymb{\Theta}$ and $\bdsymb{h}_\perp$ 
are now real (skew matrices) instead of complex (antihermitian matrices). 
This similarity is a result of the homomorphism
\EQs
\begin{pmatrix}
0 & p_\parallel & \vec{p}_{1\perp} \\ 
-p_\parallel & 0 &\vec{p}_{2\perp} \\ 
-\trans{\vec{p}_{1\perp}} & -\trans{\vec{p}_{2\perp}} & \bdsymb{p_\perp}
\end{pmatrix}
\mapsto 
\begin{pmatrix}
-\i p_\parallel & \vec{p}_{1\perp} +\i\vec{p}_{2\perp} \\ 
-\htrans{\vec{p}_{1\perp} +\i\vec{p}_{2\perp}} & \bdsymb{p_\perp}
\end{pmatrix}
\endEQs
of $\vs{so}(N+1)$ into $\vs{u}(N)$,
such that $[\vs{so}(N+1),\vs{so}(N+1)] \subset \vs{su}(N) \subset \vs{u}(N)$. 

Proceeding as before, we use equations
\eqref{sothetaeq}--\eqref{soheq} to eliminate
\EQs
\bdsymb{\Theta} &=&
\frac{1}{2}\Dinv{x}(
\bar\vvec\otimes \wvec + \vvec\otimes\bar\wvec 
-\bar\wvec\otimes \vvec -\wvec\otimes \bar\vvec ) ,
\label{sotheta}\\
\theta &=&
\frac{\i}{2}\Dinv{x}(\bar\wvec \cdot \vvec - \bar\vvec\cdot \wvec ) ,
\label{sothetatr}\\
\bdsymb{h}_{\perp} &=&
\frac{1}{2}\Dinv{x}( 
\bar\vvec\otimes\hvec_{\perp} +\vvec\otimes\bar\hvec_{\perp} 
-\bar\hvec_{\perp} \otimes\vvec - \hvec_{\perp} \otimes\bar\vvec ) ,
\\
\hpar &=&
\frac{\i}{2}\Dinv{x}(
\bar\hvec_{\perp}\cdot \vvec -\bar\vvec\cdot \hvec_{\perp} ) ,
\endEQs
in terms of the variables $\vvec$, $\hvec_{\perp}$, $\wvec$.
Then equation \eqref{soveq} gives a flow on $\vvec$,
\EQs\label{sofloweq}
\vvec_t &=&
\D{x}\wvec 
+ \frac{1}{2}\Dinv{x}(\wvec\cdot \bar\vvec-\vvec \cdot\bar\wvec)\vvec
\nonumber\\&&
+ \frac{1}{2}\vvec\hook \Dinv{x}( 
\bar\vvec\otimes \wvec + \vvec\otimes \bar\wvec 
-\bar\wvec\otimes \vvec -\wvec\otimes \bar\vvec )
-\i\hvec_{\perp}
\endEQs
with
\EQs
\wvec &=&
-\i\D{x}\hvec_{\perp}
+\frac{\i}{2}\Dinv{x}( 
\bar\hvec_{\perp}\cdot\vvec -\bar\vvec \cdot\hvec_{\perp} )\vvec 
\nonumber\\&&\qquad
+ \frac{\i}{2}\vvec \hook \Dinv{x}( 
\bar\hvec_{\perp} \otimes\vvec +\hvec_{\perp}\bar \otimes\vvec 
- \bar\vvec\otimes \hvec_{\perp} - \vvec\otimes \bar\hvec_{\perp} )
\endEQs
obtained from equation \eqref{soweq}. 
We thus read off the operators
\EQ
\Hop =
\D{x}
-\i\Dinv{x}(\bar\vvec{\cdot^\dagger \i}\ )\vvec
+\frac{1}{2}\vvec \hook \Dinv{x}( 
(\bar\vvec \wedge^\dagger\ ) - (\ \wedge^\dagger\bar\vvec) ) ,\qquad
\Iop = \i ,
\label{soHop}
\endEQ
and define the related operator
\EQ
\Jop =\Iop^{-1}\circ\Hop\circ\Iop=
\D{x}+\Dinv{x}(\bar\vvec \cdot^\dagger\ )\vvec 
+ \frac{1}{2}\vvec \hook\Dinv{x}(
(\bar\vvec \odot^\dagger\ ) - (\ \odot^\dagger\bar\vvec) )
\label{soJop}
\endEQ
using the Hermitian versions of the wedge product 
$\vec{A} \wedge^\dagger \vec{B} :=
\vec{A} \otimes \vec{B} -\bar{\vec{B}} \otimes\bar{\vec{A}}$,
the symmetric product 
$\vec{A} \odot^\dagger \vec{B} :=
\vec{A} \otimes \vec{B} +\bar{\vec{B}} \otimes\bar{\vec{A}}$,
and the dot product
$\vec{A} \cdot^\dagger \vec{B} := 
\frac{1}{2}\vec{A} \cdot\vec{B}
+\frac{1}{2}\bar{\vec{B}}\cdot\bar{\vec{A}}$,
introduced before. 

These operators $\Hop,\Jop,\Iop$ and variables $\vvec,\wvec,\hvec_\perp$
determine a very similar Hamiltonian structure 
in the space $G=SO(N+1)$ compared to $G=SU(N)$

\begin{proposition}
Theorem~\ref{thm1} and Corollary~\ref{cor1}
apply verbatim here (with the same method of proof)
to the flow equation \eqref{sofloweq}
and to the operators \eqrefs{soHop}{soJop}, 
apart from a change in the scalar curvature factor
$\chi=1$
connected with the Riemannian geometry of $SO(N+1)$.
\end{proposition}

Thus, 
\EQ
\Rop=
\i( \D{x}+\Dinv{x}(\bar\vvec \cdot^\dagger\ )\vvec 
+ \frac{1}{2}\vvec \hook\Dinv{x}(
(\bar\vvec \odot^\dagger\ ) - (\ \odot^\dagger\bar\vvec) ) )
\label{soRop}
\endEQ
yields a hereditary recursion operator
generating a hierarchy of $U(N-1)$-invariant commuting 
bi-Hamiltonian flows on $\vvec$, corresponding to 
commuting Hamiltonian vector fields
$\i\hvec^{(k)}_{\perp}$ 
and involutive covector fields
$\wvec^{(k)} = \delta H^{(k)}/\delta \bar\vvec$, 
$k=0,1,2,\ldots$.
The hierarchy starts from 
$\i\hvec^{(0)}_{\perp}=\i\vvec$, 
$\wvec^{(0)}=-\vvec$, 
which generates phase rotations,
and is followed by 
$\i\hvec^{(1)}_{\perp}=-\vvec_x$, 
$\wvec^{(1)}=-\i\vvec_x$, 
which generates $x$-translations. 
All these flows have the same recursion relations \eqref{hwhierarchy}
as in the space $G=SU(N)$,
and they also share the same NLS scaling symmetry
$x\rightarrow\lambda x$, $\vvec\rightarrow\lambda^{-1}\vvec$. 

The $+1$ and $+2$ flows given by 
$\hvec_{\perp}=\vvec$ and $\hvec_{\perp}=\i\vvec_x$ 
respectively yield 
a vector NLS equation 
\EQ\label{sonlseq}
\i\vvec_t =
\vvec_{2x}+|\vvec|^2\vvec -\frac{1}{2}\vvec\cdot\vvec \bar\vvec +\chi\vvec
\endEQ
up to a phase term (which can be absorbed by a phase rotation on $\vvec$),
and a complex vector mKdV equation 
\EQ\label{somkdveq}
\vvec_t =
\vvec_{3x}+\frac{3}{2}( 
|\vvec|^2\vvec_x +(\vvec_x\cdot\bar\vvec)\vvec -(\vvec_x\cdot\vvec)\bar\vvec )
+ \chi\vvec_x
\endEQ
up to a convective term 
(which can be absorbed by a Galilean transformation). 
Note these two equations differ compared to the ones arising
in the space $G=SU(N)$. 
The higher odd- and even- flows yield 
higher-order versions of equations \eqrefs{sonlseq}{somkdveq}. 

This hierarchy of flows corresponds to geometrical motions of the
curve $\gamma(t,x)$ obtained from equation \eqref{soet}
in a similar fashion to the ones for $G=SU(N)$ 
via identifying
$\gamma_t \leftrightarrow\tframe{}$,
$\gamma_x \leftrightarrow \xframe{}$,
$\covder{x}\gamma_x \leftrightarrow[\xconx{}{},\xframe{}]=\covD{x}\xframe{}$,
and so on as before, where
$\covder{x} \leftrightarrow \covD{x}=\D{x}+[\xconx{}{},\cdot]$.
Here we have
\EQs
[\xconx{}{},\xframe{}] &=&
\begin{pmatrix}
0 & 0 & -\vvec_2 \\ 0 & 0 & \vvec_1\\ 
\trans{\vvec_2} & -\trans{\vvec_1} & \bdsymb{0}
\end{pmatrix} ,
\\ 
\ad([\xconx{}{},\xframe{}])\xframe{} &=& 
-\ad(\xframe{})[\xconx{}{},\xframe{}] = 
\begin{pmatrix}
0 & 0 & -\vvec_1 \\ 0 & 0 & -\vvec_2\\ 
\trans{\vvec_1} & \trans{\vvec_2} & \bdsymb{0}
\end{pmatrix} ,
\endEQs
\EQs
[\xconx{}{},[\xconx{}{},\xframe{}]] &=&
\begin{pmatrix}
0 & \vvec_1{}^2 + \vvec_2{}^2 & \vec{0} \\ 
-(\vvec_1{}^2 + \vvec_2{}^2) & 0 & \vec{0} \\ 
\trans{\vec{0}} & \trans{\vec{0}} & 
2\vvec_2\otimes\vvec_1 - 2\vvec_1\otimes\vvec_2
\end{pmatrix}
\\ 
&=&
\ad([\xconx{}{},\xframe{}])^2\xframe{} ,
\nonumber
\endEQs
and so on,
where $\ad(\cdot)$ denotes the standard adjoint representation acting in the
Lie algebra $\vs{so}(N+1)$.

The $+1$ flow
\EQ
\hvec_{\perp}=\vvec ,\qquad
\hpar= 0 ,\qquad
\bdsymb{h}_{\perp}= \bdsymb{0} ,
\endEQ
gives the frame equation
\EQ
\tframe{} =
\begin{pmatrix}
0 & 0 & \vvec_1 \\ 0 & 0 & \vvec_2\\ 
-\trans{\vvec_1} & -\trans{\vvec_2} & \bdsymb{0}
\end{pmatrix}
=\ad(\xframe{})[\xconx{}{},\xframe{}] ,
\label{sonlsframeflow}
\endEQ
so thus $\gamma(t,x)$ satisfies 
the chiral Schr\"odinger map equation \eqref{nlsmap}
on the Lie group $G=SO(N+1)$.
All higher odd-flows on $\vvec$ in the hierarchy 
determine higher-order chiral Schr\"odinger map equations. 

Next, the $+2$ flow
\EQ
\hvec_{\perp}=\i\vvec_x ,\qquad
\hpar= \frac{1}{2}|\vvec|^2 ,\qquad
\bdsymb{h}_{\perp}=
\frac{\i}{2}( \bar\vvec \otimes \vvec - \vvec \otimes \bar\vvec )
\endEQ
yields the frame equation
\EQs
\tframe{} &=&
\begin{pmatrix}
0 & \frac{1}{2}(\vvec_1{}^2 + \vvec_2{}^2) & -(\vvec_2)_x \\ 
-\frac{1}{2}(\vvec_1{}^2 + \vvec_2{}^2) & 0 & (\vvec_1)_x \\ 
\trans{(\vvec_2)_x} & -\trans{(\vvec_1)_x} & 
\vvec_2\otimes\vvec_1 - \vvec_1\otimes\vvec_2
\end{pmatrix}
\\ \nonumber &=&
\D{x}[\xconx{}{},\xframe{}]
-\frac{1}{2}[\xconx{}{},[\xconx{}{},\xframe{}]] 
\nonumber
\endEQs
which gives the same frame equation as in the space $G=SU(N)$, 
\EQ
\tframe{} =
\covD{x}[\xconx{}{},\xframe{}]
-\frac{3}{2}\chi^{-1} \ad([\xconx{}{},\xframe{}])^2\xframe{}
\endEQ
up to the change in the scalar curvature factor, $\chi=1$. 
Thus, 
$\gamma(t,x)$ satisfies the chiral mKdV map equation \eqref{mkdvmap}
on the Lie group $G=SO(N+1)$.
All higher even- flows on $\vvec$ in the hierarchy 
yield higher-order chiral mKdV map equations for $\gamma$.

These same geometric nonlinear PDEs were found to arise \cite{kievpaper}
from curve flows in 
the corresponding symmetric space $G/SO(N) \simeq S^N$. 

The hierarchy also contains a $-1$ flow in which 
$\hvec_{\perp}$ is annihilated by
the symplectic operator $\Jop$
so it lies in the kernel $\Rop(\hvec_{\perp})=0$
of the recursion operator.
This flow has the same geometrical meaning 
as in the space $G=SU(N)$, namely 
$\Jop(\hvec_{\perp})=\wvec=0$
whence $\tconx{}{}=0$
which implies 
$0=[\tconx{}{},\xframe{}]=\covD{t}\xframe{}$
where
$\covD{t} = \D{t}+ [\tconx{}{},\cdot]$.
Thus, the correspondence 
$\covder{t} \leftrightarrow \covD{t}$, 
$\gamma_x \leftrightarrow \xframe{}$ 
directly yields the chiral wave map equation \eqref{wavemap}
on the Lie group $G=SO(N+1)$. 
The resulting $-1$ flow equation on $\vvec$ is given by 
\EQ\label{sovsgflow}
\vvec_t=-\chi\i \hvec_{\perp} ,\qquad
\chi=1
\endEQ
where $\hvec_{\perp}$ satisfies the equation
\EQ\label{sowsgflow}
0 = \i\wvec =\D{x} \hvec_{\perp} + \i\hpar\vvec +\vvec \hook\bdsymb{h}_\perp
\endEQ
together with equations \eqrefs{sohpareq}{soheq}. 
Similarly to the case $G=SU(N)$, 
these three equations determine 
$\hvec_{\perp}$, $\hpar$, $\bdsymb{h}_\perp$
as nonlocal functions of $\vvec$ (and its $x$ derivatives).
Proceeding as before, we seek 
an inverse local expression for $\vvec$ 
obtained through an algebraic reduction
\EQ\label{sohsgeq}
\bdsymb{h}_\perp = 
\alpha\i( \bar\hvec_{\perp} \otimes\hvec_{\perp} 
-\hvec_{\perp} \otimes\bar\hvec_{\perp} )
\endEQ
for some expression $\alpha(\hpar)\in\Rnum{}$.
Substitution of $\bdsymb{h}_\perp$ into equation \eqref{soheq}, 
followed by the use of equations \eqrefs{sohpareq}{sowsgflow}, 
gives
\EQ\label{soalphaeq}
\alpha = -\frac{1}{2} \hpar{}^{-1} .
\endEQ
We next use the wave map conservation law \eqref{wavemapconslaw}
where now 
$|\gamma_t|_{g}^{2}$\break
$= <\tframe{},\tframe{}>_\vs{p}
= |\hvec_{\perp}|^2 +\hpar{}^2 +\frac{1}{2}|\bdsymb{h}|^2$,
corresponding to the conservation law 
\EQ
0 =\D{x}( |\hvec_{\perp}|^2 +\hpar{}^2 +\frac{1}{2}|\bdsymb{h}_\perp|^2 )
\endEQ
admitted by equations \eqref{sowsgflow}, \eqref{sohpareq}, \eqref{soheq}
with
\EQ\label{sotrheq}
|\bdsymb{h}_\perp|^2 := -\tr(\bdsymb{h}_\perp{}^2)
= 2\alpha^2( |\hvec_{\perp}|^4 -\hvec_\perp{}^2 \bar\hvec_\perp{}^2 ) 
\endEQ
where
$\hvec_\perp{}^2 := \hvec_\perp\cdot \hvec_\perp$, 
$\bar\hvec_\perp{}^2 := \bar\hvec_\perp\cdot\bar\hvec_\perp$. 
As before, 
a conformal scaling of $t$ can be used to make
$|\gamma_t|_{g}$ equal to a constant.
By putting $|\gamma_t|_{g}=1$ we obtain 
\EQ
1 =|\hvec_{\perp}|^4 +\hpar{}^2 +\frac{1}{2}|\bdsymb{h}_\perp|^2 
= \frac{1}{4}\hpar{}^{-2}( 
|\hvec_{\perp}|^4 -\hvec_\perp{}^2 \bar\hvec_\perp{}^2 ) 
+\hpar{}^2 +|\hvec_{\perp}|^2 
\endEQ
from equations \eqrefs{soalphaeq}{sotrheq}.
This yields a quadratic equation
\EQ\label{sosghpareq}
0= \hpar{}^4 +(|\hvec_{\perp}|^2 -1)\hpar{}^2
+ |\hvec_{\perp}|^4 -\hvec_\perp{}^2 \bar\hvec_\perp{}^2
\endEQ
determining 
\EQs
2 \hpar{}^2 
&=&
1-|\hvec_\perp|^2 \pm 
\sqrt{1-2|\hvec_{\perp}|^2+ \hvec_\perp{}^2 \bar\hvec_\perp{}^2} ,
\label{sosghpar}\\
2 \alpha^2 
&=&
( |\hvec_{\perp}|^4 -\hvec_\perp{}^2 \bar\hvec_\perp{}^2 )^{-1}
\Big( 1-|\hvec_\perp|^2 \mp 
\sqrt{1-2|\hvec_{\perp}|^2+ \hvec_\perp{}^2 \bar\hvec_\perp{}^2} \Big) .
\nonumber\\
\label{sosgalpha}
\endEQs
The flow equation \eqref{sovsgflow} allows these variables to be expressed 
in terms of $\vvec$:
\EQ
|\hvec_{\perp}|^2 = |\vvec_t|^2 ,\qquad
\hvec_\perp{}^2 =\vvec_t{}^2 ,\qquad
\bar\hvec_\perp{}^2 = \bar\vvec_t{}^2 .
\endEQ
Similarly, equation \eqref{sowsgflow} 
combined with equations \eqrefs{sohsgeq}{soalphaeq}
yields
\EQ\label{sovhsgeq}
\hvec_{\perp} = 
\i\Dinv{x}( \frac{1}{2}\alpha^{-1} \vvec 
+\alpha( (\vvec\cdot\vvec_t) \bar\vvec_t -(\vvec\cdot\bar\vvec_t) \vvec_t ) ) .
\endEQ

Thus the $-1$ flow equation on $\vvec$ becomes
the nonlocal evolution equation 
\EQ
\vvec_t = 
\frac{1}{\sqrt{2}}\Dinv{x}( 
B_{\mp}\vvec +|B|^{-1} B_{\pm} ( (\vvec\cdot\bar\vvec_t) \vvec_t 
-(\vvec\cdot\vvec_t) \bar\vvec_t ) )
\endEQ
where
\EQs
B_{\pm}^2 &:=& 
1 -|\vvec_t|^2\pm \sqrt{1-2|\vvec_t|^2 -\vvec_t{}^2\bar\vvec_t{}^2} 
=|B|^2 /B_{\mp}^2 ,
\\
|B|^2 &:=& 
|\vvec_t|^4 -\vvec_t{}^2 \bar\vvec_t{}^2 =B_+^2 B_-^2 .
\endEQs
The hyperbolic form of this vector PDE 
is a complex variant of a vector SG equation
(with a factor $\sqrt{2}$ absorbed into a scaling of $t$)
\EQ\label{sosghyperboliceq}
\vvec_{tx}=
B_{\mp}\vvec +|B|^{-1} B_{\pm} ( (\vvec\cdot\vvec_t) \bar\vvec_t 
-(\vvec\cdot\bar\vvec_t) \vvec_t )
\endEQ
which was found in \cite{AncoWolf}. 

There is an equivalent hyperbolic equation on $\hvec_{\perp}$ 
given by an inverse for expression \eqref{sovhsgeq} as follows.
Substitution of equation \eqref{sohsgeq} into equation \eqref{sowsgflow}
first yields the relation 
\EQ
\i\vvec = 
2\alpha( \hvec_{\perp}{}_x +\i\alpha( 
(\vvec\cdot\bar\hvec_{\perp}) \hvec_{\perp}
- (\vvec\cdot\hvec_{\perp}) \bar\hvec_{\perp} ) ) . 
\endEQ
Then by taking its dot product separately with 
$\hvec_{\perp}$ and $\bar\hvec_{\perp}$, 
we obtain the additional relations
\EQs
&&
\i (1-|\hvec_{\perp}|^2 -2\hpar{}^2) \vvec\cdot\hvec_\perp
= 
(\alpha |\hvec_{\perp}|^2 -\frac{1}{2}\alpha^{-1}) 
\hvec_\perp\cdot\hvec_{\perp x} 
- \alpha \hvec_\perp{}^2 \bar\hvec_\perp\cdot\hvec_{\perp}{}_x ,
\nonumber\\
&&
\i (1-|\hvec_{\perp}|^2 -2\hpar{}^2) \vvec\cdot\bar\hvec_\perp
= 
\alpha \hvec_\perp{}^2 \bar\hvec_\perp\cdot\hvec_{\perp x} 
-(\alpha |\hvec_{\perp}|^2 +\frac{1}{2}\alpha^{-1}) 
\hvec_\perp\cdot\hvec_{\perp}{}_x , 
\nonumber
\endEQs
which thus determines 
$\vvec\cdot\hvec_\perp$ and $\vvec\cdot\bar\hvec_\perp$
and hence $\vvec$ 
in terms of $\hvec_\perp$, $\bar\hvec_\perp$, and $\hvec_\perp{}_x$.
Finally, 
substitution of these expressions into the flow equation \eqref{sovsgflow}
yields the complex vector SG equation
\EQs
\hvec_{\perp} &=&
\Big( \alpha\Big( 2\hvec_{\perp}{}_x 
\mp (1-2|\hvec_{\perp}|^2+ \hvec_\perp{}^2 \bar\hvec_\perp{}^2)^{-1/2}(
\nonumber\\&&\quad
( (1-2\alpha^2 |\hvec_{\perp}|^2) \hvec_{\perp}\cdot \hvec_{\perp}{}_x 
+2\alpha^2 \hvec_{\perp}{}^2 \bar\hvec_{\perp}\cdot \hvec_{\perp}{}_x )
\bar\hvec_{\perp}
\nonumber\\&&\quad
-( (1+2\alpha^2 |\hvec_{\perp}|^2) \bar\hvec_{\perp}\cdot \hvec_{\perp}{}_x 
+2\alpha^2 \bar\hvec_{\perp}{}^2 \hvec_{\perp}\cdot \hvec_{\perp}{}_x )
\hvec_{\perp} ) 
\Big)\Big){}_t .
\label{soSGeq}
\endEQs

Note that, as written, 
the hyperbolic PDEs \eqrefs{sosghyperboliceq}{soSGeq}
for $G=SO(N+1)$
are valid only when $|B|\neq 0$,
which holds precisely in the vector case, $N>2$. 
The scalar case $N=2$ becomes a singular limit $|B|=0$,
as seen from the quadratic equation \eqref{sosghpareq} 
whose solutions \eqref{sosghpar} degenerate to 
$\hpar{}^2 = \frac{1}{2} B_\pm^2 = 1-|\hvec_\perp|^2,0$
in the $+/-$ cases respectively. 
Thus $\alpha =-\frac{1}{2}\hpar{}^{-1}$ is well-defined only in the $+$ case,
with 
\EQ
\hpar{}^2 = \frac{1}{2} B_+^2 = 1-|\hvec_\perp|^2 ,\qquad N=2 ,
\endEQ
where we have the corresponding limit 
\EQ
\alpha^2 =\lim_{|B|\rightarrow 0} \frac{1}{2} |B|^{-2} B_-^2 
= \frac{1}{4}(1-|\hvec_\perp|^2)^{-1} ,\qquad N=2 ,
\endEQ
and where we identify 
the $1$-component complex vectors $\hvec_\perp,\vvec,\wvec\in \Cnum{}$ 
with complex scalars. 
(This settles the questions raised in \cite{AncoWolf} concerning the
existence of a scalar limit for the hyperbolic vector equation 
\eqref{sosghyperboliceq}.)

In this limit the hyperbolic PDEs \eqrefs{sosghyperboliceq}{soSGeq}
for $G=SO(3)$ 
reduce to the scalar case of 
the hyperbolic PDEs \eqrefs{susghyperboliceq}{suSGeq}
for $G=SU(2)$,
due to the local isomorphism of the Lie groups $SU(2)\simeq SO(3)$. 
The same happens for the evolutionary PDEs in the hierarchies 
for $G=SO(3)$ and $G=SU(2)$, namely 
the scalar case of the NLS equations \eqrefs{sonlseq}{sunlseq}
and the mKdV equations \eqrefs{somkdveq}{sumkdveq} 
each coincide (up to scalings of the variables). 

The symmetry-integrability classification results 
in \cite{AncoWolf} 
show that the hyperbolic vector equation \eqref{sosghyperboliceq}
admits the vector NLS equation \eqref{sonlseq} as a higher symmetry.
We see that, from Corollary~\ref{cor1} 
applied to the recursion operator \eqref{soRop}, 
there is 
a hierarchy of vector NLS/mKdV higher symmetries
\EQs
\vvec_{t}^{(0)} &=&
\i\vvec ,
\label{so0flow}\\
\vvec_{t}^{(1)} &=& \Rop(\i\vvec) 
= -\vvec_x ,
\label{so1flow}\\
\vvec_{t}^{(2)} &=& \Rop^2(\i\vvec) =
-\i(\vvec_{2x}+|\vvec|^2\vvec -\frac{1}{2}\vvec\cdot\vvec \bar\vvec ) ,
\label{so2flow}\\
\vvec_{t}^{(3)} &=& \Rop^2(-\vvec_x) =
\vvec_{3x}+\frac{3}{2}( 
|\vvec|^2\vvec_x +(\vvec_x\cdot\bar\vvec)\vvec 
-(\vvec_x\cdot\vvec)\bar\vvec ) ,
\label{so3flow}
\endEQs
and so on,
generated by this operator $\Rop$,
while the adjoint operator $\Rop^*$ generates 
a corresponding hierarchy of NLS/mKdV Hamiltonians
(modulo total derivatives)
\EQs
H^{(0)} &=&
|\vvec|^2 ,
\nonumber\\
H^{(1)} &=&
\i\vvec_x\cdot\bar\vvec ,
\nonumber\\
H^{(2)} &=&
-|\vvec_x|^2 +\frac{1}{2}|\vvec|^2 -\frac{1}{4}\vvec^2\bar\vvec^2 ,
\nonumber\\
H^{(3)} &=&
\i\vvec_x\cdot(\bar\vvec_{2x} +\frac{3}{2}|\vvec|^2\bar\vvec)
-\i\frac{3}{8}(\vvec^2)_x \bar\vvec^2 ,
\nonumber
\endEQs
and so on.
All of Hamiltonians are conserved densities for the $-1$ flow 
\EQ\label{so-1flow}
\vvec_{t}^{(-1)} = -\i\hvec_{\perp}
\endEQ
associated to the vector SG equation \eqref{soSGeq},
and hence they determine a hierarchy of conservation laws admitted 
for the hyperbolic vector equation \eqref{sosghyperboliceq}.
Likewise all of the symmetries comprise a hierarchy that commutes
with the $-1$ flow 
and are therefore admitted symmetries of 
the hyperbolic vector equation \eqref{sosghyperboliceq}. 

All of the vector PDEs \eqsref{so0flow}{so3flow}, etc.,
viewed as flows on $\vvec$, 
including the $-1$ flow \eqref{so-1flow}, 
possess the NLS scaling symmetry
$x\rightarrow\lambda x$, $\vvec\rightarrow\lambda^{-1}\vvec$,
with $t\rightarrow\lambda^{k} t$
for $k=-1,0,1,2,\ldots$, 
where these PDEs for $k\geq 0$ will be local polynomials in the variables
$\vvec,\vvec_x,\vvec_{2x},\ldots$
in the same manner as before. 

\begin{theorem}\label{thm4}
In the Lie group $SO(N+1)$
there is a hierarchy of bi-Hamiltonian flows of curves $\gamma(t,x)$
described by geometric map equations.
The $0$ flow is a convective (traveling wave) map \eqref{convmap},
while the $+1$ flow is a non-stretching 
chiral Schr\"odinger map \eqref{nlsmap}
and the $+2$ flow is a non-stretching 
chiral mKdV map \eqref{mkdvmap},
and the other odd- and even- flows are higher order analogs.
The kernel of the recursion operator \eqref{soRop} in the hierarchy
yields the $-1$ flow which is a non-stretching chiral wave map \eqref{wavemap}.
Moreover the components of the principal normal vector along the
$+1,+2,-1$ flows in a $SU(N)$-parallel frame respectively satisfy 
a vector NLS equation \eqref{sonlseq}, 
a complex vector mKdV equation \eqref{somkdveq}
and a complex vector hyperbolic equation \eqref{sosghyperboliceq}. 
\end{theorem}

\section{ Concluding remarks }

The Lie groups $SO(N+1)$ and $SU(N)$ each contain 
a hierarchy of integrable bi-Hamiltonian curve flows described by 
a chiral Schr\"odinger map equation \eqref{nlsmap} for the $+1$ flow, 
a chiral mKdV map equation \eqref{mkdvmap} for the $+2$ flow, 
and a chiral wave map equation \eqref{wavemap} for the $-1$ flow
coming from the kernel of the recursion operator of each hierarchy. 
The principal normal components in a parallel frame along these flows
in each Lie group 
satisfy $U(N-1)$-invariant soliton equations respectively given by 
a vector NLS equation, a complex vector mKdV equation,
and a hyperbolic vector equation related to a complex vector SG equation.

These two Lie groups are singled out as exhausting the isometry groups $G$
that arise for compact Riemannian symmetric spaces of the type $G/SO(N)$
as known from Cartan's classification \cite{Helgason}. 
Moreover, since $G=SO(N+1)$ is locally isomorphic to $G=SU(N)$ 
when (and only when) $N=2$,
the integrable hierarchies of curve flows in the spaces 
$SO(3) \simeq SU(2) \simeq S^3$ therefore coincide 
precisely in the scalar case, 
with the $+1,+2,-1$ flows reducing to 
$U(1)$-invariant scalar soliton equations
consisting of the NLS equation 
$\i v_t=v_{2x} +2 |v|^2 v$
and complex versions of mKdV and SG equations
$v_t =v_{3x} +6|v|^2 v_x$ 
and 
$v_{tx} = 2\sqrt{1-|v_t|^2}) v$
(up to rescalings of $v$ and $t$). 

The present results thus account for the existence of the two 
unitarily-invariant vector generalizations of 
the NLS equation and the complex mKdV and SG equations
that are known from symmetry-integrability classifications 
\cite{SokolovWolf,AncoWolf}. 
Moreover, their bi-Hamiltonian integrability structure
as summarized by the operators $\Rop=\Hop\circ\Iop=\Iop\circ\Jop$
is shown to be geometrically encoded in the frame structure equations
for the corresponding curve flows in the two Lie groups 
$G=SO(N+1),SU(N) \subset U(N)$. 
This encoding utilizes a parallel moving frame formulation
based on earlier work \cite{kievpaper} studying
integrable curve flows in the Riemannian symmetric spaces $G/SO(N)$. 
Indeed, 
the bi-Hamiltonian operator structure derived in \cite{kievpaper} 
for curve flows in $G/SO(N)$ can be recovered from 
$\Hop$ and $\Jop$ if the connection variables $\vvec$ and $\wvec$
are restricted to be real while the flow-direction variable $\hvec_\perp$
is restricted to be imaginary, 
in which case the $-1$ flow and all even-flows reduce to
the hierarchy of flows in $G/SO(N)$.
More particularly, the operator $\Rop^2 =-\Hop\circ\Jop$ 
acts as a vector NLS/mKdV recursion operator which is (up to a sign) 
a complex version of the vector mKdV recursion operator 
coming from $G/SO(N)$. 

Finally, there is a broad generalization \cite{forthcoming} of these results
yielding hierarchies of group-invariant soliton equations
associated to integrable curve flows described by geometric map equations
in all semisimple Lie groups and Riemannian symmetric spaces.

\section*{Acknowledgments}
S.C.A is supported by an N.S.E.R.C. grant.

\end{document}